\documentclass{IEEEojcsys}

\usepackage[colorlinks,urlcolor=blue,linkcolor=blue,citecolor=blue]{hyperref}

\usepackage{color,array}
\usepackage{graphicx}
\let\labelindent\relax

\usepackage[font={small,bf,sf}]{caption}
\usepackage[utf8]{inputenc}
\usepackage{amsmath}
\usepackage{amssymb}
\usepackage{amsthm}
\usepackage{xcolor}
\usepackage{enumitem}
\usepackage{fixltx2e}
\usepackage{caption}
\usepackage{stfloats}
\usepackage[most]{tcolorbox}
\usepackage{subcaption}
\usepackage{cleveref}
\newtcolorbox{mymathbox}[1][]{colback=white, sharp corners, #1}

\Crefname{figure}{\text{Fig.}}{\text{Figs.}}
\Crefname{equation}{}{}

\newcommand{\B}[1]{\if#1\relax\bm{#1}\else\mathbf{#1}\fi} % bold text
\newcommand{\R}{\mathbb{R}}
\newcommand{\sL}{\mathcal{L}}
\newcommand{\sN}{\mathcal{N}}
\newcommand{\sKL}{\mathcal{KL}}
\newcommand{\sK}{\mathcal{K}}
\newcommand{\abs}[1]{\vert #1 \vert}
\newcommand{\norm}[1]{\Vert #1 \Vert}
\newcommand{\black}[1]{{\color{black} #1}}

\newcommand{\T}{^{\mathsf{T}}}

\newtheorem{defn}{Definition}

\newtheorem{lem}{Lemma}
\newtheorem{cor}{Corollary}
\newtheorem{rmk}{Remark}
\newtheorem{prop}{Proposition}

\jvol{00}
\jnum{XX}
\paper{1234567}
\pubyear{2021}
\receiveddate{XX September 2021}
\accepteddate{XX October 2021}
\publisheddate{XX November 2021}
\currentdate{XX November 2021}
\doiinfo{OJCSYS.2021.Doi Number}

\begin{document}

\sptitle{Article Category}

\title{A Multiplex Approach Against Disturbance Propagation in Nonlinear Networks with Delays\footnote{asda}}

\editor{This paper was recommended by Associate Editor F. A. Author.}

\author{Shihao Xie\affilmark{1} }

\author{Giovanni Russo\affilmark{2}  (Senior Member, IEEE)}

\affil{School of Electrical and Electronic Engineering, University College Dublin, Dublin, Ireland} 
\affil{Department of Information and Electrical Engineering and Applied Mathematics, University of Salerno, Salerno, Italy} 

\corresp{CORRESPONDING AUTHOR: Giovanni Russo (e-mail: \href{mailto: giovarusso@unisa.it}{giovarusso@unisa.it})}
\authornote{An early version of this paper, see \cite{https://doi.org/10.48550/arxiv.2202.07638}, without
proofs and considering a special case only, is presented at NecSys 22.}
\markboth{PREPARATION OF PAPERS FOR IEEE OPEN JOURNAL OF CONTROL SYSTEMS}{Shihao Xie {\itshape ET AL}.}
\begin{abstract}
We consider both leaderless and leader-follower, possibly nonlinear, networks affected by time-varying communication delays. For such systems, we give a set of sufficient conditions that guarantee the convergence of the network towards some desired behaviour while simultaneously ensuring the rejection of polynomial disturbances and the non-amplification of other classes of disturbances across the network. To fulfill these desired properties, and prove our main results, we propose the use of a control protocol that implements a multiplex architecture. The use of our results for control protocol design is then illustrated in the context of formation control. The protocols are validated both in-silico and via an experimental set-up with real robots. All experiments confirm the effectiveness of our approach. 
\end{abstract}

\begin{IEEEkeywords}
Nonlinear systems and control, Large-scale systems, Multiplex networks, Time delays, Disturbance propagation
% Enter key words or phrases in alphabetical order, separated by commas. For a list of\goodbreak suggested keywords, send a blank e-mail to \href{mailto:keywords@ieee.org}{keywords@ieee.org} or visit\goodbreak \href{http://www.ieee.org/organizations/pubs/ani_prod/keywrd98.txt}{http://www.ieee.org/organizations/pubs/ani\_prod/keywrd98.txt}
\end{IEEEkeywords}

\maketitle

\section{INTRODUCTION}\label{sec:introduction}
Driven by the introduction of low-cost, high performance and connected devices, network systems have considerably increased their size and complexity. 
In this context, a key challenge is that of designing networks that not only fulfil some desired behaviour, but also: (i) reject certain classes of disturbances; (ii) do not amplify across the network disturbances that are not rejected. {These properties can be captured via a {\em scalability} property of the network (see Section III-\ref{sec:control_goal} for the rigorous definition) which denotes the preservation of desired 
properties uniformly with respect to the number of agents.} We present a number of sufficient conditions that guarantee these properties for nonlinear network systems with delays. {The conditions are based on multiplex architectures,}
%Our conditions lead to a multiplex network architecture 
that we exploit for control design, {and on tools from contraction theory}. 
\subsection{RELATED WORKS} 
We briefly survey key related works on network scalability, disturbance rejection in networks and contraction theory.

{\paragraph*{Network scalability and disturbance rejection}}
The study of how disturbances propagate within a network system is a central topic for the platooning of autonomous vehicles. In particular, the key idea behind several definitions of string stability in literature is that of giving upper bounds on the deviations induced by disturbances that are uniform with respect to platoon size, see e.g. {\cite{MONTEIL2019198}, where the bounds are found by leveraging contraction theory and \cite{STUDLI2017157} for a survey that includes approaches based on passivity arguments}. 
% These works assume delay-free inter-vehicle communications.
% and an extension to delayed platoons with linear vehicle dynamics can be found in e.g. \cite{6891349}. 
% In this context, we also recall \cite{https://doi.org/10.48550/arxiv.2209.08993} where networks with heterogeneous delays are considered and network stability (a weaker notion than scalability) is investigated in the presence of disturbances. 
% Consensus stability for (disturbance-free) robotic networks subject to delays is instead considered in \cite{9601198}. 
{In these works, no delays are considered and agents are arranged along a string.} For networks with {more general topologies and with} delay-free interconnections, we recall results on mesh stability \cite{mesh} and on leader-to-formation stability, which is considered in \cite{1303690} and it characterizes network behaviour with respect to inputs from leaders. For delay-free leaderless networks with regular topology, a scalability property has been recently investigated in \cite{8370724}, where Lyapunov-based conditions are given; we also recall \cite{10156367} that studies scalability of delayed-free networks and \cite{tegling2023scale} that investigates scalable stability of consensus algorithm for delay-free networks under a wide range of graphs. For networks with arbitrary topologies and {with} delays, sufficient conditions for scalability can be found in \cite{9353260}, which leverages contraction theory arguments for time-delayed systems. {In this last work, non-amplification of disturbances is guaranteed. However, results in \cite{9353260} do not guarantee disturbance rejection. Among different classes of disturbances, of particular interest is to design controllers able to reject polynomial disturbances. To this aim, disturbance observers can be designed to estimate, and compensate, these disturbances \cite{5457987, 6425973}. However, these works do not guarantee non-amplification as these disturbances are spread across the network.} {A complementary approach to compensate polynomial disturbances consists in leveraging distributed integral actions. By pursuing this approach,} the problem of constant disturbances rejection is tackled in the context of string stability for delay-free platoons \cite{knorn2014passivity, SILVA2021109542}. 
{Finally, in this work we present a hardware validation of the protocol designed in accordance with our results. In this context, we recall \cite{di2015design} that studies platoon of  vehicles with linear dynamics and delayed communications and the designed control is validated experimentally on three vehicles. In \cite{li2020longitudinal}, a platoon of third order integrators without communication delays is considered and validation is performed via hardware experiments with $4$ cars. In \cite{hu2023spontaneous}, a distributed guiding vector-field algorithm is designed for a group of robots modelled by single integrator delay-free dynamics to achieve string stability and its effectiveness is validated on hardware experiments with unmanned surface vessels. In all these works, however, no disturbance rejection is guaranteed.}

{\paragraph*{Contraction theory}}
Contracting systems exhibit transient and asymptotic behaviors \cite{LOHMILLER1998683} that are desirable when designing network systems \cite{centorrino2023euclidean}; we refer to \cite{aminzare2014contraction, tsukamoto2021contraction} and references therein for further details. We also recall \cite{1618853} which shows, using Euclidean metric, how contraction is preserved through certain time-delayed communications and \cite{8561231} where conditions for the synthesis of distributed controls are given by using separable metric structures. In this context, we also recall \cite{COOGAN2019349} where separable Lyapunov functions are constructed for monotone systems that are also contractive. For delay-free systems, based on the use of contraction, a sufficient condition for stability of a feedback loop consisting of an exponentially stable multi-input multi-output nonlinear plant and an integral controller (to compensate constant disturbances) has been obtained in \cite{9247460}. The problem of constant output regulation for a class of input-affine multi-input multi-output nonlinear systems with constant disturbances, has also been recently tackled via contraction in \cite{9627575}. Other works have also shown that contraction using non-Euclidean metrics can be useful to study a wide range of biological~\cite{entrainment_2010}, neural~\cite{Marg_22,jafarpour2021robust} and engineered~\cite{MONTEIL2019198} networks. We also recall~\cite{1531-3492_2021188} the recent extension of contraction to dynamical systems on time scales, i.e. systems evolving on arbitrary (potentially non-uniform) time domains.

{\paragraph*{Summary}
The stream of works surveyed above highlights that, currently, no results are available to design protocols that simultaneously guarantee, for nonlinear networks with delays and, possibly, leaders providing time-varying references: (i) the fulfilment of a desired behaviour for the network; (ii) rejection of polynomial disturbances; (iii) non-amplification of other classes of disturbances. Protocols guaranteeing these key requirements for network systems can instead be designed with our results.}

\subsection{STATEMENT OF CONTRIBUTIONS}
We give sufficient conditions for the exponential convergence of possibly nonlinear network systems with communication delays towards some desired behaviour, while guaranteeing rejection of polynomial disturbances and non-amplification of other classes of disturbances. {To the best of our knowledge, these are the first conditions that allow to guarantee these desired properties.} Specifically, our contribution can be summarised as follows:
\begin{itemize}
    \item[{\bf (i)}] we formalize these desired properties for the network with the notions of $\sL_\infty^p$-Input-to-State Scalability and $\sL_\infty^p$-Input-Output Scalability;
    \item[{\bf (ii)}] we introduce a set of sufficient conditions for scalability. The conditions, {based on the use of a multiplex  architecture}, leverage non-Euclidean contraction arguments and certain structured norms. This allows to consider both leaderless and leader-follower networks of nonlinear, possibly heterogeneous agents with communication delays. {Moreover,} this approach also allows to consider arbitrary network topologies and time-varying references. {We are not aware of other results that allow to design protocols guaranteeing these properties};
    \item[{\bf (iii)}] we leverage our results to design control protocols for formation control problems, allowing a formation to track a time-varying reference provided by a leader, reject polynomial disturbances and ensure the non-amplification of other classes of disturbances. {We also show how the fulfilment of the sufficient conditions can be recast as an optimization problem;}
    %Also, our conditions can be conveniently recast into a convex optimisation problem that allows to design the control protocol; %% NOT SO SURE IF WE SHOULD SAY THIS %%
    \item[{\bf (iv)}] finally, we validate our protocols both {\em in-silico} and {via a multi-robot formation control application with} real robots. All the experiments confirm the effectiveness of our approach\footnote{Documented code and data to replicate all our results are available at \url{http://tinyurl.com/46xvfy7f} together with recordings of the experiments.} {with the protocols effectively guaranteeing that the robots achieve, and track, the desired formation while simultaneously guaranteeing rejection of polynomial disturbances and the non-amplification of other classes of disturbances.}
    %\blue{To the best of our knowledge, there is no real hardware implementation in the literature of string stability or network scalability that validates the designed control protocol guaranteeing non-amplification of disturbances in the presence of communication delays.}
\end{itemize}
To the best of our knowledge, these are the first results that, for nonlinear networks with delays, simultaneously guarantee tracking of some desired behaviour, rejection of polynomial disturbances and non-amplification of other disturbances. In fact, our results directly extend \cite{MONTEIL2019198,SILVA2021109542}, which are focused on string stability of platoon systems, and our prior work \cite{9353260, https://doi.org/10.48550/arxiv.2202.07638}. Specifically: (i) in \cite{MONTEIL2019198} string stability was considered for platoons without delays and polynomial disturbances rejection is not guaranteed; (ii) in \cite{SILVA2021109542} constant disturbances (i.e., zero order polynomials) are rejected under the assumption that the platoon is delay-free; (iii) in \cite{9353260} delays are instead considered but disturbance rejection is not guaranteed {(indeed, as also illustrated in Section IV-\ref{sec: application}, the protocols designed with the results from this paper tackle situations that cannot be considered with \cite{9353260})}; {(iv) in \cite{https://doi.org/10.48550/arxiv.2202.07638} only first order disturbance rejection is considered and no proof is given.} Moreover, while in \cite{https://doi.org/10.48550/arxiv.2209.08993} polynomial disturbances are rejected for networks affected by heterogeneous delays, the non-amplification of other disturbances across the network is not guaranteed.

{The rest of the paper is organised as follows. In Section \ref{sec: preliminary}, we give mathematical preliminaries necessary for the development of the main results of the paper. In Section \ref{sec:problem_set-up}, we give the set-up, introducing both the multiplex architecture and the notion of scalability together with the control problem. The main theoretical results are then given in Section \ref{sec: main_results} and validated on a robot formation application via both simulations and hardware experiments in Section \ref{sec:validation}. Concluding remarks are given in Section \ref{sec:conclusions}.}

\section{MATHEMATICAL PRELIMINARIES}\label{sec: preliminary}
Let $A$ be a $m \times m$ real matrix, we denote by $\norm{A}_p$ the matrix norm induced by $p$-vector norm $\abs{\cdot}_p$. The matrix measure of $A$ induced by $\abs{\cdot}_p$ is  $\mu_p(A):=\lim_{h\rightarrow 0^+}\frac{\norm{I+hA}_p-1}{h}$. We write $A \succeq 0$ when $A$ is positive semi-definite and $A \preceq 0$ when it is negative semi-definite. The symmetric part of $A$ is  $[A]_s:=\frac{A+A\T}{2}$.
Given a piece-wise continuous signal $w_i(t)$, we let $\norm{w_i(\cdot)}_{\sL_\infty^p}:=\sup_t\abs{w_i(t)}_p$. We denote by $I_n$ the $n\times n$ identity matrix and by $0_{m\times n}$ the $m\times n$ zero matrix (if $m=n$ we simply write $0_n$). The Kronecker product is denoted by $\otimes$. For a generic set $\mathcal{A}$, its cardinality is denoted by $\abs{\mathcal{A}}$. Let $f$ be a  smooth function, we denote by $f^{(n)}$ the $n$-th derivative of $f$. The Dini derivative of a continuous function $g$ is denoted as $D^+g(x):=\limsup_{h\rightarrow 0+}\frac{g(x+h)-g(x)}{h}$. Given a vector $\eta:=[\eta_1\T,\ldots,\eta_N\T]\T$, $\eta_i \in \mathbb{R}^n$, we denote the structured vector norm $\abs{\cdot}_G$ as $\abs{\eta}_G := \abs{\left[ \abs{\eta_1}_{G_1},\ldots,\abs{\eta_N}_{G_N}\right]}_S$ with $\abs{\cdot}_{G_i}$ being norms on $\R^n$ and $\abs{\cdot}_{S}$ being norms on $\R^N$. The matrix norm and matrix measure induced by $\abs{\cdot}_G$ ($\abs{\cdot}_{G_i}$) are denoted by $\norm{\cdot}_G$, $\mu_G(\cdot)$ ($\norm{\cdot}_{G_i}, \mu_{G_i}(\cdot)$), respectively.

We recall that a continuous function $\alpha: [0,a)\rightarrow [0,\infty)$ is said to belong to class $\sK$ if it is strictly increasing and $\alpha(0)=0$. It is said to belong to class $\sK_\infty$ if $a=\infty$ and $\alpha(r)\rightarrow \infty$ as $r \rightarrow \infty$. A continuous function $\beta: [0,a)\times [0,\infty)\rightarrow [0,\infty)$ is said to belong to class $\sKL$ if, for each fixed $s$, the mapping $\beta(r,s)$ belongs to class $\sK$ with respect to $r$ and, for each fixed $r$, the mapping $\beta(r,s)$ is decreasing with respect to $s$ and $\beta(r,s)\rightarrow 0$ as $s\rightarrow \infty$.

We let $\abs{\cdot}_S$ and $\mu_S(\cdot)$ be, respectively, any $p$-vector norm and its induced matrix measure on $\mathbb{R}^N$. In particular, the norm $\abs{\cdot}_S$ is monotone, i.e. for any non-negative $N$-dimensional vector $x,y \in \R_{\ge 0}^N$, $x\le y$ implies that $\abs{x}_S \le \abs{y}_S$ where the inequality $x\le y$ is component-wise. {Given a matrix $A\in \mathbb{R}^{nN \times nN}$, we partition it into:
\begin{align*}
    A = \left[\begin{matrix}A_{11} & A_{12} & \ldots & A_{1N}\\A_{21} & A_{22} & \ldots & A_{2N}\\
    \vdots & \vdots & \ddots & \vdots\\
    A_{N1} & A_{N2} & \ldots & A_{NN}
    \end{matrix}\right]
\end{align*}
We define the following: 
\begin{align*}
\begin{split}
    \hat{A}_{ii}=\mu_{G_i}(A_{ii}), \ \hat{A}_{ij}=\norm{A_{ij}}_{G_{i,j}},
\end{split}
\end{align*}
where $\norm{A_{ij}}_{G_{i,j}}:=\sup_{\abs{x}_{G_i}=1}\abs{A_{ij}x}_{G_j}$ and we also define 
\begin{align*}
\begin{split}
    \bar{A}_{ii}=\norm{A_{ii}}_{G_{i,i}},\ \bar{A}_{ij}=\norm{A_{ij}}_{G_{i,j}}.
\end{split}
\end{align*}
Finally, we define
\begin{align*}
    \hat{A} = \left[\begin{matrix}\hat A_{11} & \hat A_{12} & \ldots & \hat A_{1N}\\\hat A_{21} & \hat A_{22} & \ldots & \hat A_{2N}\\
    \vdots & \vdots & \ddots & \vdots\\
    \hat A_{N1} & \hat A_{N2} & \ldots & \hat A_{NN}
    \end{matrix}\right]
\end{align*}
and
\begin{align*}
    \bar{A} = \left[\begin{matrix}\bar A_{11} & \bar A_{12} & \ldots & \bar A_{1N}\\ \bar A_{21} & \bar A_{22} & \ldots & \bar A_{2N}\\
    \vdots & \vdots & \ddots & \vdots\\
    \bar A_{N1} & \bar A_{N2} & \ldots & \bar A_{NN}
    \end{matrix}\right]
\end{align*}
Then we can state the following lemma which follows \cite{5717887}. }
\begin{lem}\label{lem:matrix norm}
{For any structured vector norm $\abs{\cdot}_G$ on $\R^{nN\times nN}$ and any $p$-vector norm $\abs{\cdot}_S$ on $\R^N$},  
% \begin{enumerate}
% \item $A:=({A_{ij}})_{i,j=1}^N \in \mathbb{R}^{nN \times nN}$, $A_{ij} \in \mathbb{R}^{n \times n}$; 
% \item $\hat{A}:=(\hat{A}_{ij})_{i,j=1}^N \in \mathbb{R}^{N \times {N}}$, with $\hat{A}_{ii}:=\mu_{G_i}(A_{ii})$ and $\hat{A}_{ij}:=\norm{A_{ij}}_{G_{i,j}}$, {$\norm{A_{ij}}_{G_{i,j}}:=\sup_{\abs{x}_{G_i}=1}\abs{A_{ij}x}_{G_j}$};
% \item $\bar{A}:=(\bar{A}_{ij})_{i,j=1}^N \in \mathbb{R}^{N \times N}$, with $\bar{A}_{ij}:=\norm{A_{ij}}_{G_{i,j}}$. 
% \end{enumerate}
{we have}: \\
(i) $\mu_G(A) \le \mu_S(\hat{A})$;
(ii)$\norm{A}_G \le \norm{{\bar A}}_S$.  
\end{lem}
The next lemma is adapted from \cite[Theorem $2.4$]{WEN2008169}.
\begin{lem}\label{prop:halanay}
Let $u:[t_0-\tau_{\max},+\infty)\rightarrow\R_{\ge 0}$ , $0\le\tau_{\max}<+\infty$. {If the following inequality}
$$D^+u(t) \le au(t)+b \sup_{t-\tau_{\max} \le s \le t}u(s)+c, \ \  t\ge t_0,$$
{holds} with: 
\begin{itemize}
    \item [(1)]  $u(t)=\abs{\varphi(t)}$, $\forall t\in [t_0-\tau_{\max},t_0]$ where $\varphi(t)$ is bounded in $[t_0-\tau_{\max},t_0]$; 
    \item[(2)] $a < 0$, $b \ge 0$ and $c \ge 0$ and that there exists some $\sigma>0$
such that $a+b \le -\sigma <0, \forall t\ge t_0$.
\end{itemize} 
Then:
$$
u(t) \le \sup_{t_0-\tau_{\max} \le s \le t_0}u(s)e^{- \lambda (t-t_0)}+\frac{c}{\sigma},
$$
where $\lambda>0$ is the solution of 
$\lambda+ a+be^{\lambda\tau_{\max}}=0.$
\end{lem}
% \blue{Consider a dynamical system
% \begin{equation}\label{equ: contraction}
%     \dot{x}=f(t,x(t))
% \end{equation}
% with $f: \R_{\ge 0}\times \R^n\rightarrow \R^n$. The following definition is classic in contraction related literature, see e.g. \cite{entrainment_2010}.
% \begin{defn}
% Given a matrix norm $\norm{\cdot}$, the associated matrix measure $\mu(\cdot)$, and a constant $c>0$ $(c=0)$, \eqref{equ: contraction} is said to be strongly (weakly) infinitesimally contracting if
% \begin{align*}
%     \mu\left(\frac{\partial f(t,x)}{\partial x}\right)\le -c
% \end{align*}
% \end{defn}

% }

\section{THE SET-UP}\label{sec:problem_set-up}
% We now describe the proposed control architecture (Section \Cref{sec:architecture}) and formalize the control goal (Section \Cref{sec:control_goal}).
We consider a network system of $N>1$ agents with the dynamics of the $i$-th agent given by
\begin{align}\label{equ: dynamics}
\begin{split}
    \dot{x}_i(t)&=f_i(x_i,t)+u_i(t)+d_i(t), \ \ t\ge t_0 \ge 0,\\
    y_i(t)&=g_i(x_i),
\end{split}
\end{align}
with initial conditions $x_i(t_0)$, $i=1,\dots,N$, and where: (i) $x_i(t)\in \R^n$ is the state of the $i$-th agent; (ii) $u_i(t) \in \R^n$ is the control input; (iii) $d_i(t) \in \R^n$ is an external disturbance signal on the agent; (iv) $f_i: \R^n\times \R_{\ge 0} \rightarrow \R^n$ is the intrinsic dynamics of the agent, which is assumed to be smooth; (v) $g_i:\R^n\rightarrow\R^q$ is the output function for the $i$-th agent. We consider disturbances of the form: 
\begin{align}\label{equ: disturbance}
    d_i(t)=w_i(t) + \bar{d}_i(t) := w_i(t)+\sum_{k=0}^{m-1} \bar{d}_{i,k}\cdot t^k,
\end{align}
where $w_i(t)$ is a piece-wise continuous signal and $\bar{d}_{i,k}$'s are constant vectors. Disturbances in \Cref{equ: disturbance} embeds polynomial disturbance $\bar{d}_i(t)$, of order $m-1$, and $w_i(t)$ that captures the {\em residual} terms in the disturbance that are not polynomial {(we consider a rather general class of signals for $w_i(t)$ as we only require that this is piece-wise continuous)}. {Polynomial disturbances are widely used in literature and embed disturbances that are typically used for control design, i.e., constant disturbances, ramp disturbances etc. The disturbances \Cref{equ: disturbance} include several interesting} special cases. {For example, consider the case} where: (i) $m=1$ in \Cref{equ: disturbance}. Then, we have $d_i(t) = w_i(t) + \bar{d}_{i,0}$. In the context of platooning, these types of disturbances model situations when a platoon of vehicles encounters a slope where $\bar{d}_{i,0}$ models constant disturbance on vehicle acceleration and $w_i(t)$ models {disturbances caused by} {\em small bumps} along the slope \cite{SILVA2021109542}; {(ii) $m=2$, $\bar{d}_{i,0}=0$, $w_i(t)=0$ in \Cref{equ: disturbance} we have $d_i(t) = \bar{d}_{i,1}\cdot t$. In the context of power systems, this ramp disturbance can model an attack to the system \cite{6740883}.}

% For example, in a special case where $m=1$ we have $d_i(t) = w_i(t) + \bar{d}_{i,0}$.  As noted in e.g. \cite{SILVA2021109542}, in the context of platooning, these types of disturbances model situations when a platoon of vehicles encounters a slope: this gives rise to a constant disturbance on the vehicle acceleration and the term $w_i(t)$ can then model {\em small bumps} along the slope. 

\begin{rmk}\label{rmk:disturbance}
%We consider disturbances consisting of a polynomial component and a piece-wise continuous component. 
Polynomial disturbances are commonly considered in the literature. See e.g. \cite{5457987} where observers for these disturbances are devised and \cite{6425973} where the problem of rejecting these disturbances is considered. The signal $w_i(t)$ can be physically interpreted as a (typically, {\em small}) {\em discrepancy} between the polynomial disturbance model and the actual disturbance signal. For platooning, rejection of constant disturbances (i.e. the disturbance in \Cref{equ: disturbance} when $m=1$) has been considered in \cite{knorn2014passivity, SILVA2021109542}.
\end{rmk}

\subsection{MULTIPLEX ARCHITECTURE}\label{sec:architecture}
\begin{figure}[thbp]
\centering
\includegraphics[width=0.8\linewidth]{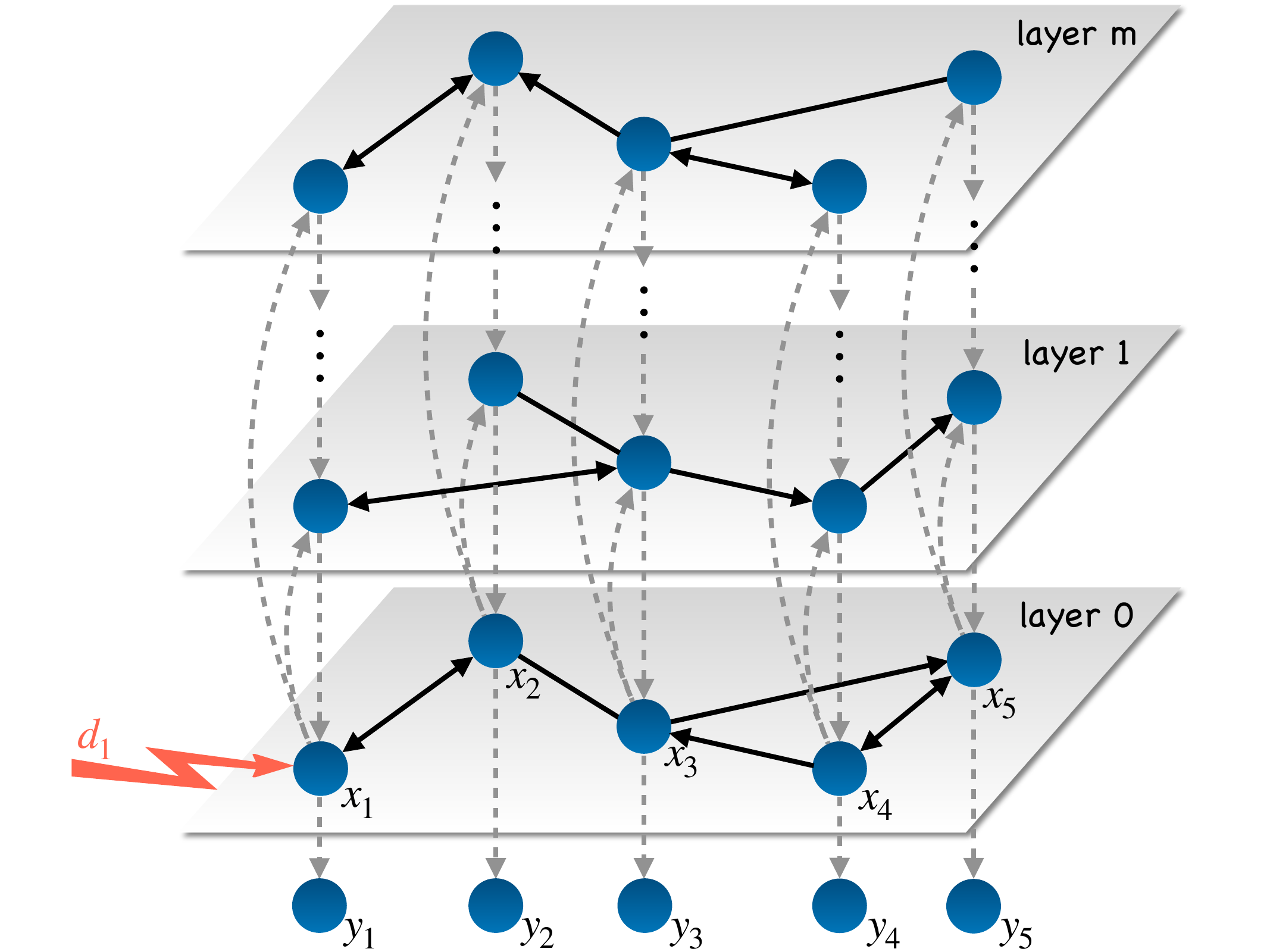}
\caption{The multiplex architecture. One disturbance is highlighted and the reference signal is omitted. Layers can have different topologies, which can be both directed and undirected.}
\label{fig:multiplex_network}
\end{figure}
As we shall see, with our main results, we give sufficient conditions guaranteeing: (i) tracking of a desired reference; (ii) rejection of the $\bar{d}_i(t)$'s in \Cref{equ: disturbance}; (iii) non-amplification of the $w_i(t)$'s, which do not need to be polynomials for our results to hold. These properties are all captured by the notion of scalability (see Section III-\ref{sec:control_goal} for the rigorous definition).
%See Section \Cref{sec:control_goal} for the formulation of the control goal. 
%Given the network system \Cref{equ: dynamics} affected by disturbance \Cref{equ: disturbance}, a desired property is to ensure the fulfillment of some desired behaviour while simultaneously guaranteeing that the polynomial disturbance $\bar{d}_i(t)$ is rejected and the residual disturbance $w_i(t)$ is not amplified -- this is captured by the notion of scalability. See Section \Cref{sec:control_goal} for the rigorous definition of scalability. 
To fulfill such a property, we propose the use of the multiplex architecture schematically shown in \Cref{fig:multiplex_network}. In such a figure, the network system is in layer $0$ and the multiplex layers (i.e. layer $1,\ldots,m$) concur to build up the control protocol. This is of the form:
\begin{align}\label{equ: control}
    \begin{split}
        u_i(t) &=h_{i,0}(x_i,\{x_j\}_{j\in\sN_i},x_l,t)+h_{i,0}^{(\tau)}(x_i,\{x_j\}_{j\in\sN_i},x_l,t)\\
        &+r_{i,1}(t),\\
        \dot{r}_{i,1}(t)&=h_{i,1}(x_i,\{x_j\}_{j\in\sN_i},x_l,t)
        +h_{i,1}^{(\tau)}(x_i,\{x_j\}_{j\in\sN_i},x_l,t)\\
        &+r_{i,2}(t),\\
        & \ \vdots \\
        \dot{r}_{i,m}(t)&=h_{i,m}(x_i,\{x_j\}_{j\in\sN_i},x_l,t)+h_{i,m}^{(\tau)}(x_i,\{x_j\}_{j\in\sN_i},x_l,t),
    \end{split}
\end{align}
where $r_{i,k}(t)$ is the output generated by the multiplex layer $k \in \{1,\ldots,m\}$. As illustrated in \Cref{fig:multiplex_network}, the multiplex layer $k \in \{1,\ldots,m\}$ receives information from the agents (on layer $0$) and outputs a signal to the layer immediately below, i.e. layer $k-1$. In \Cref{equ: control}: (i) $\{x_j\}_{j\in\mathcal{N}_i} \in \R^{nN_i}$ denotes the stack of the states of the neighbours of agent $i$ where $\sN_i$ is the set of neighbours of agent $i$ and $N_i:=\abs{\sN_i}$ is the cardinality of the set which is assumed to be bounded $\forall N$. That is, $\forall N$, there exists some $\bar{N}<\infty$ such that  $N_i\le \bar{N}$, $\forall i$; (ii) $x_l(t):=[x_{l_1}\T(t),\ldots,x_{l_M}\T(t)]\T$ is the reference signal, possibly provided by a group of $M$ leaders; (iii) {$\forall k\in\{0,\ldots,m\}$ we have $h_{i,k}: \R^{n} \times \R^{nN_i} \times \R^{nM} \times \R_{\ge 0} \rightarrow \R^{n}$ and $h_{i,k}^{(\tau)}: \R^{n} \times \R^{nN_i} \times \R^{nM} \times \R_{\ge 0} \rightarrow \R^{n}$} are smooth functions that model the {\em delay-free} and {\em delayed} couplings from neighbours and (possible) leaders with bounded time-varying delay $\tau(t)$, satisfying $\tau(t)\le \tau_{\max}$, respectively. {In \eqref{equ: control}, $\tau(t)$ is the same for all agents. We use the term homogeneous delay for cases where all delayed couplings are affected by the same delay. This assumption is commonly seen in literature, see \Cref{rmk: homogeneous_delay}.} Note that not all the agents necessarily receive information from leaders (if any). {Situations where there is an overlap between delay-free and delayed communication naturally occur in various applications. For instance, in the context of platooning, certain states, such as radar-based separation data from nearby vehicles, may be readily available to an agent without any significant delay. On the other hand, the information of separation from more distant vehicles or control actions of neighbouring vehicles may require communication and be subject to delays caused by measurement or processing. In a special case where the delay-free couplings are all $0$, i.e. $h_{i,k}(x_i,\{x_j\}_{j\in\sN_i},x_l,t)=0, \forall k$, \eqref{equ: control} can also model the situation when there are only delayed couplings.} Without loss of generality, in \Cref{equ: control} we set, $\forall s \in [t_0-\tau_{\max}, t_0], \forall i=1,\ldots,N$, $\forall k=1,\ldots, m$,  $x_i(s)=\varphi_i(s)$ and $r_{i,k}(s)=\phi_{i,k}(s)$, with $\varphi_i(s)$ and $\phi_{i,k}(s)$ being continuous and bounded functions in $ [t_0-\tau_{\max},t_0]$. 
\begin{rmk}\label{rmk: homogeneous_delay}
{The results rely on the fact that delays are homogeneous. This setup naturally arises when some {\em time-stamping} is available for state information exchanged between nodes. Homogeneous delays in fact arise when the state information is transmitted at a sequence of time points, i.e. $\{\ldots,t_{k-1}, t_{k},\ldots \}$, the information available at time $t\in (t_{k-1},t_k)$ is then $x(t_{k-1}):=x(t-\tau(t))$ where $\tau(t)=t-t_{k-1}$ \cite{fridman2010refined}. In a broader context,} network systems affected by these homogeneous delays naturally arise in the context of multi-agent systems. For example, in \cite{9462542} string stability of platoon is studied when agents are affected by homogeneous constant delays. Homogeneous delays are also considered in the context of consensus \cite{doi:10.1137/110832574} and synchronization \cite{guo2018event}. 
\end{rmk}
{\begin{rmk}
The multiplex architecture implements a distributed integral action. As we shall see, if the architecture is designed in accordance with our result, this can be used to both reject high order polynomial disturbances and to guarantee the non-amplification of the residual disturbance. We derive such sufficient conditions later in Section III-\ref{sec: technical}.
\end{rmk}}

\begin{rmk}
We do not require the delay $\tau(t)$ to be known explicitly. We only make the rather standard assumption, see e.g. \cite{doi:10.1137/110832574, guo2018event} and references therein, that $\tau(t)$ is bounded by $\tau_{\max}$, which as well does not need to be known.
\end{rmk}

\begin{rmk}
We do not require that the multiplex layers have the same topology. This degree of freedom in the design of the layers can be leveraged, as noted in \cite[Remark 12]{BURBANOLOMBANA2016310}, to e.g. reduce the {\em control interventions} across the network.
\end{rmk}
% \begin{rmk}\label{rmk:control_examples}
% Protocols of the form of (\Cref{equ: control}) naturally arise in a wide range of applications, where agents might have access to a mixture of delay-free and delayed information. Typical examples can be found in the context of e.g. vehicle platooning \cite{6891349}, formation control \cite{9353260}. Moreover, the couplings are commonly of diffusive type in these control protocols.
% % Choices for the coupling functions in \Cref{equ: control} are typically of diffusive type, e.g. $h_{i,0}(x_i,\{x_j\}_{j\in\sN_i},x_l,t)=k_i\sum_{l_i\in l}(x_{l_i}(t) - x_i(t)-\delta_{l_i i}^\ast)$ and $h_{i,0}^{(\tau)}(x_i,\{x_j\}_{j\in\sN_i},x_l,t)=k_i^{(\tau)}\sum_{j\in\mathcal{N}_i}(x_j(t-\tau(t)) - x_i(t-\tau(t)) -\delta_{ji}^\ast)$, where $k_i, k_i^{(\tau)}$ are the coupling strength and $\delta_{l_i i}^\ast$, $\delta_{ji}^\ast$ denotes the desired offset from leader $l_i$, agent $j$ to agent $i$, respectively.
% \end{rmk}
\subsection{NETWORK SCALABILITY AND CONTROL GOAL}\label{sec:control_goal}
We let $x(t)=[x_1\T(t), \ldots, x_N\T(t)]\T$ be the stack of the agent states, $u(t)=[u_1\T(t),\ldots,u_N\T(t)]\T$ be the stack of the control inputs, $d(t)=[d_1\T(t),\ldots,d_N\T(t)]\T$ be the stack of the disturbances, $w(t)=[w_1\T(t),\ldots,w_N\T(t)]\T$ be the stack of the residual disturbances, $\bar{d}(t)=[\bar d_1\T(t),\ldots,\bar d_N\T(t)]\T$ be the stack of the polynomial disturbances {and $r_{i}(t) = [r_{i,1}\T(t),\ldots,r_{i,m}\T(t)]\T$}. 
%Network scalability is expressed in terms of the so-called desired solution of the disturbance-free (unperturbed in what follows) system. 
In what follows, we say that  $[x_1^{\ast\mathsf{T}}(t), r_1^{\ast\mathsf{T}}(t), \ldots,x_N^{\ast\mathsf{T}}(t), r_N^{\ast\mathsf{T}}(t)]\T$ is the desired solution for a network system \Cref{equ: dynamics} controlled by \Cref{equ: control} when $d_i(t) = 0$, $\forall i$, if: (i) $\dot{x}_i^\ast(t)=f_i(x_i^\ast(t),t)$ with $x_i^\ast(s)=x_i^\ast(t_0), s\in[t_0-\tau_{\max},t_0]$; (ii) $r_{i,k}^{\ast}(t) = 0$, $\forall i$, $\forall k$ and $\forall t$. In what follows, we simply say that $x^\ast(t)=[x_1^{\ast\mathsf{T}}(t),\ldots,x_N^{\ast\mathsf{T}}(t)]\T$ is the desired solution of \Cref{equ: dynamics}, leaving it implicit that $r_{i,k}^\ast(t) =0$, $\forall i$, $\forall k$ and $\forall t$.
%{This is the solution of the system when there are no disturbances characterized by having: (i) the state of the agents keeping some desired configuration; (ii) the multiplex layers giving no contribution to the $u_i(t)$'s. That is, $[x_1^{\ast\mathsf{T}}(t), r_1^{\ast\mathsf{T}}(t), \ldots,x_N^{\ast\mathsf{T}}(t), r_N^{\ast\mathsf{T}}(t)]\T$ is the desired solution for a network system \Cref{equ: dynamics} controlled by \Cref{equ: control} when $d_i(t) = 0$, $\forall i$, if: (i) $\dot{x}_i^\ast(t)=f_i(x_i^\ast(t),t)$ with $x_i^\ast(s)=x_i^\ast(t_0), s\in[t_0-\tau_{\max},t_0]$; (ii) $r_{i,k}^{\ast}(t) = 0$, $\forall i$, $\forall k$ and $\forall t$. In what follows, we simply say that $x^\ast(t)=[x_1^{\ast\mathsf{T}}(t),\ldots,x_N^{\ast\mathsf{T}}(t)]\T$ is the desired solution of \Cref{equ: dynamics}, leaving it implicit that $r_{i,k}^\ast(t) =0$, $\forall i$, $\forall k$ and $\forall t$.} 
In the special case where the closed-loop system has: (i) no multiplex layers, this notion of desired solution yields the one used in \cite{9353260,MONTEIL2019198,8370724} to formalize their control goal;  (ii) one multiplex layer, such a notion yields the desired solution used in \cite{SILVA2021109542} to characterize string stability. The desired output of \Cref{equ: dynamics} is  $y^\ast(t):=[y_1^{*\mathsf{T}}(t), \dots,y_N^{\ast\mathsf{T}}(t)]\T$, with ${y}^\ast_i(t)=g_i(x_i^\ast(t)), \forall i$. 

We are now ready to introduce the notion of {\em scalability}. For the closed-loop network \Cref{equ: dynamics} - \Cref{equ: control}, scalability implies the fulfilment of the following properties simultaneously: (i) tracking of $x^\ast(t)$; (ii) rejection of $\bar{d}(t)$; (iii) non-amplification of $w(t)$ across the nodes. {The following definition extends the notion of Disturbance String Stability (DSS) in \cite{SILVA2021109542} and $\mathcal{L}_\infty$-scalable-Input-to-State Stability/$\mathcal{L}_\infty$-scalable-Input-Output Stability ($\mathcal{L}_\infty$-sISS/$\mathcal{L}_\infty$-sIOS) in our previous work \cite{9353260} by taking into account the effects of both delays and polynomial disturbances in the upper bounds. In fact, DSS does not consider delays and only takes into account the effects of the constant disturbance while $\mathcal{L}_\infty$-sISS/$\mathcal{L}_\infty$-sIOS does not consider the effect of polynomial disturbances.} 
% This makes scalability a stronger property than stability as the need of fulfilling (iii) is not required to achieve stability. 
\begin{defn}\label{def: L_inf}
Consider the closed-loop system \Cref{equ: dynamics} - \Cref{equ: control} with disturbance $d(t)=w(t) + \bar{d}(t)$. The system is
\begin{itemize}
    \item $\sL_\infty^p$-Input-to-State Scalable with respect to $w(t)$: if there exists class $\sKL$ functions $\alpha(\cdot,\cdot)$, $\beta(\cdot,\cdot)$, a class $\sK$ function $\gamma(\cdot)$, such that for any initial condition and $\forall t \ge t_0$, 
    \begin{align*}
        \begin{split}
            &\max_i\abs{x_i(t)-x_i^\ast(t)}_p \le \gamma\left(\max_i\norm{w_i(\cdot)}_{\sL_\infty^p}\right) + \\
            &\alpha\left(\max_i\sup_{t_0-\tau_{\max}\le s \le t_0}\abs{x_i(s)-x_i^\ast(s)}_p,t-t_0\right) + \\ 
            &\beta\bigg(
            \max_i\sup_{t_0-\tau_{\max}\le s \le t_0}\sum_{k=1}^m\abs{r_{i,k}(s)+\bar{d}_i^{(k-1)}(s)}_p,t-t_0\bigg) 
        \end{split}
    \end{align*}
    holds $\forall N$;
    \item $\sL_\infty^p$-Input-Output Scalable with respect to $w(t)$: if there exists class $\sKL$ functions $\alpha(\cdot,\cdot)$, $\beta(\cdot,\cdot)$, a class $\sK$ function $\gamma(\cdot)$, such that for any initial condition and $\forall t \ge t_0$, 
    \begin{align*}
        \begin{split}
            &\max_i\abs{y_i(t)-y_i^\ast(t)}_p\le \gamma\left(\max_i\norm{w_i(\cdot)}_{\sL_\infty^p}\right)+\\
            &\alpha\left(\max_i\sup_{t_0-\tau_{\max}\le s \le t_0}\abs{x_i(s)-x_i^\ast(s)}_p,t-t_0\right)+\\ 
            &\beta\bigg(\max_i\sup_{t_0-\tau_{\max}\le s \le t_0}\sum_{k=1}^m\abs{r_{i,k}(s)+\bar{d}_i^{(k-1)}(s)}_p,t-t_0\bigg)
        \end{split}
    \end{align*}
    holds $\forall N$.
\end{itemize}
\end{defn}
{\Cref{def: L_inf} gives upper bounds for state/output deviation composed of the norm of: (i) initial state deviation; (ii) time derivatives of polynomial disturbance plus the output from multiplex layers; and (iii) the residual disturbance. Note also that the bounds in \Cref{def: L_inf} are uniform in the number of agents, $N$. Hence, scalability is a stronger property than stability, which does not require uniformity of the bounds in $N$.} It in turn guarantees that residual disturbances are not amplified within the network system. In the special case when: (i) $\bar{d}(t) =0$ and there are no multiplex layers, \Cref{def: L_inf} becomes the {$\mathcal{L}_\infty$-sISS} in \cite{9353260}; {and (ii) $\bar{d}(t)=[\bar{d}_{1,0},\ldots, \bar{d}_{N,0}]\T$ and there is one multiplex layer, \Cref{def: L_inf} becomes DSS in \cite{SILVA2021109542}}. In what follows, whenever it is clear from the context we simply say that the network is $\sL_\infty^p$-Input-to-State Scalable ($\sL_\infty^p$-Input-Output Scalable) if \Cref{def: L_inf} is fulfilled. In the special case where $p=2$ we simply say that the network is $\sL_\infty$-Input-to-State Scalable ($\sL_\infty$-Input-Output Scalable).

{Given the set-up of this section, we can now formulate our control objective. Specifically, given the network system \Cref{equ: dynamics}, our goal is to design control protocols of the form of \eqref{equ: control} so that the closed-loop system fulfils \Cref{def: L_inf}.}

% In the next section, we give a running example of the control problem we consider which is the multi-robot formation control.}

\subsection{RUNNING EXAMPLE: MULTI-ROBOT FORMATION CONTROL}
\begin{figure}%[thbp]
\centering
\includegraphics[width=\linewidth]{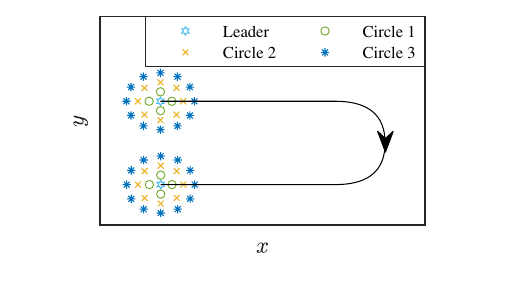}
\caption{Reference trajectory of the hand position provided by the virtual leader together with an example of desired formation.}
\label{fig:formation_pattern}
\end{figure}

We consider formation control for a group of $N$ unicycle robots and we now introduce the set-up, illustrating the concepts introduced so far. We consider the dynamics for the robots hand position, which is described by (see e.g. \cite{wilson2022robotarium,1261347} and references therein):
\begin{align}\label{equ: fbl_dynamics}
	\dot{\eta}_i(t)={\left[\begin{matrix} \cos\theta_i(t) & -L_i\sin\theta_i(t) \\ \sin\theta_i(t) & L_i\cos\theta_i(t)\end{matrix}\right]}u_i(t)+d_i(t),
\end{align}
where $\eta_i(t)$ denotes the hand position of the $i$-th robot, $L_i\in \R_{>0}$ is the distance of the hand position to the wheel axis and $\theta_i(t)$ is the heading angle. In the above equation, $d_i(t)$ is the disturbance on the $i$-th robot and $u_i(t)$ is the control input.  For concreteness, within this running example, we consider the case where the disturbances affecting the robots, i.e. $d_i(t)=[d_i^x(t),d_i^y(t)]\T$, are of the form  
\begin{align*}
    d_i^x(t):=w_i^x(t) + \bar{d}_{i,0}^{x}+\bar{d}_{i,1}^{x}\cdot t,\\
    d_i^y(t):=w_i^y(t) + \bar{d}_{i,0}^{y}+\bar{d}_{i,1}^{y}\cdot t.
\end{align*}
These disturbances naturally arise in the context of e.g. unicycle-like marine robots whose dynamics are also captured by the above dynamics. For these robots, the constant terms in $d_i(t)$ model the disturbances due to the ocean current \cite{5979747} and the piece-wise continuous residual disturbances $w_i^x(t)$, $w_i^y(t)$ model e.g. transient variations of the current. The ramps in the disturbance can model ramp attack signals \cite{6740883}.
The dynamics in \Cref{equ: fbl_dynamics} can be feedback linearised by
\begin{align*}
    u_i(t)={\left[\begin{matrix} \cos\theta_i(t) & -L_i\sin\theta_i(t) \\ \sin\theta_i(t) & L_i\cos\theta_i(t)\end{matrix}\right]^{-1}}\nu_i(t),
\end{align*}
yielding the dynamics for the closed-loop network system
\begin{align}\label{equ: dynamics_hand_position}
	\dot{\eta}_i(t)=\nu_i(t)+d_i(t), \forall i.
\end{align}
{with 
\begin{align*}
    \nu_i(t)=\left[\begin{matrix} \cos\theta_i(t) & -L_i\sin\theta_i(t) \\ \sin\theta_i(t) & L_i\cos\theta_i(t)\end{matrix}\right]u_i(t)
\end{align*}
\noindent where $\nu_i(t)$ is the control input to be designed for the feedback linearised system \eqref{equ: dynamics_hand_position}.}
In what follows we make use of the compact notation $w_i(t):=[w_i^x(t),w_i^y(t)]\T$ and 
\begin{align}\label{equ: d_bar}
\bar{d}_i(t):=\left[\begin{matrix}\bar{d}_{i,0}^{x}+\bar{d}_{i,1}^{x}\cdot t\\
    \bar{d}_{i,0}^{y}+\bar{d}_{i,1}^{y}\cdot t\end{matrix}\right].
\end{align}
{We consider the setting of e.g., \cite{li2013leader} so that robots receive: (i) broadcast signals from a delay-free {\em virtual leader}; (ii) position information from their neighbours with a bounded time-varying delay $\tau(t)\le\tau_{\max}$. We denote by $\eta_l(t)$ and $v_l(t)$ the hand position and reference velocity from the virtual leader.} We seek to design a control protocol for the network so that the following requirements are satisfied: 
\begin{itemize}
    \item[R1] robots follow a reference trajectory and track {$v_l(t)$};
    \item[R2] desired offsets from the leader ($\delta_{li}^\ast$) and from neighbours ($\delta_{ji}^\ast$) are kept;
    \item[R3] the polynomial disturbances $\bar{d}_i(t)$ are rejected;
    \item[R4] the $w_i(t)$'s are not amplified.
\end{itemize}
These properties can be achieved if the closed-loop network \Cref{equ: dynamics_hand_position} is $\sL_\infty$-Input-to-State Scalable with the desired solution defined as $\eta^\ast(t):=[\eta_1^{\ast\mathsf{T}}(t), \ldots, \eta_N^{\ast\mathsf{T}}(t)]\T$, with  $\dot{\eta}_i^\ast(t)=v_l(t)$, $\forall i$ and  $\eta_l(t)-\eta_i^\ast(t)=\delta_{li}^\ast$, $\eta_j^\ast(t)-\eta_i^\ast(t)=\delta_{ji}^\ast$. See \Cref{fig:formation_pattern} for an example of desired robotic formation for a network with robots in $3$ concentric circles. 
In the next part of the example, given the set-up illustrated so far, we show that the $\sL_\infty$-Input-to-State Scalability property of such robotic network can be guaranteed if the protocol is designed in accordance with our main results (introduced next).

\section{MAIN METHODOLOGICAL RESULTS}\label{sec: main_results}
{We now introduce our main results to assess the scalability property given in Definition \ref{def: L_inf}. Specifically, with \Cref{prop: weighted}} we give a set of sufficient conditions for $\sL_\infty^p$-Input-to-State Scalability of the closed-loop system \Cref{equ: dynamics} - \Cref{equ: control} affected by disturbances of the form \Cref{equ: disturbance}. {With \Cref{cor:scalability} we instead give a sufficient condition for $\sL_\infty^p$-Input-Output Scalability of the system. We also continue the running example, employing the conditions for protocol design.}

\subsection{SUFFICIENT CONDITIONS FOR SCALABILITY}\label{sec: technical}
The results are stated in terms of the block diagonal matrix $T:=I_N \otimes \bar{T} \in \R^{N\cdot n \cdot (m+1) \times N\cdot n \cdot (m+1)}$ with
\begin{align}\label{equ: blk_diag}
    \bar{T}:=\left[\begin{matrix}I_{n} & \alpha_{1}\cdot I_{n} &  &  & \\
 & I_{n} & \ddots &  & \\
 &  & \ddots &  & \alpha_{m}\cdot I_{n}\\
 &  &  &  & I_{n} \end{matrix}\right] \in \R^{n\cdot (m+1) \times n\cdot (m+1)},
\end{align}
where $\alpha_{k} \in \R, \forall k \in \{1, \ldots, m\}$, is independent on $N$.
% ; \blue{and (ii) the structured norm $\abs{x}_G := \left\vert\abs{x_1}_p,\ldots,\abs{x_N}_p \right\vert_{\infty}$.}
%\GR{The structure norm does not appear at all in the statement of Proposition 1!}

\begin{prop}\label{prop: weighted}
Consider the closed-loop network system \Cref{equ: dynamics} - \Cref{equ: control} with $  y_i(t) \allowbreak  = x_i(t)$ affected by disturbances \Cref{equ: disturbance}. 
{If} $\forall t\ge t_0$, the following conditions are satisfied for some $0\le\underline{\sigma}<\bar{\sigma}<+\infty$:
\begin{figure*}[b]
\hrulefill
\begin{equation}\label{equ: matrix}
\begin{split}
    &\bar{A}_{ii}(t)=\left[\begin{matrix} \frac{\partial f_i(x_i,t)}{\partial x_i}+\frac{\partial 				h_{i,0}(x,x_l,t)}{\partial x_i} & I_n & 0_n & \cdots & 0_n\\
     			\frac{\partial h_{i,1}(x,x_l,t)}{\partial x_i} & 0_n & I_n & \cdots & 0_n\\ 						\vdots & \vdots & \vdots & \ddots & \vdots\\
     			\frac{\partial h_{i,m-1}(x,x_l,t)}{\partial x_i} & 0_n & 0_n & \cdots & I_n\\
      			\frac{\partial h_{i,m}(x,x_l,t)}{\partial x_i} & 0_n & 0_n & \cdots & 0_n 	\end{matrix}\right] \\
    \bar{A}_{ij}(t)=&\left[\begin{matrix} \frac{\partial h_{i,0}(x,x_l,t)}{\partial x_j} & 0_n 					& \cdots & 0_n\\
      			\vdots & \vdots & \ddots & \vdots\\
      			\frac{\partial h_{i,m}(x,x_l,t)}{\partial x_j} & 0_n & \cdots & 0_n 	\end{matrix}\right], \ 
    \bar{B}_{ij}(t)=\left[\begin{matrix} \frac{\partial h_{i,0}^{(\tau)}(x,x_l,t)}{\partial x_j} & 					0_n & \cdots & 0_n\\
      			\vdots & \vdots & \ddots & \vdots\\
      			\frac{\partial h_{i,m}^{(\tau)}(x,x_l,t)}{\partial x_j} & 0_n & \cdots & 0_n  \end{matrix}\right]
\end{split}
\end{equation}
\end{figure*}
\begin{itemize}
    \item[C1] $h_{i,k}(x_i^\ast, \{x_j^\ast\}_{j\in\sN_i} ,x_l,t)=h_{i,k}^{(\tau)}(x_i^\ast, \{x_j^\ast\}_{j\in\sN_i} ,x_l,t)=0$, $\forall i, \forall k$;
    \item[C2] $\mu_p(\bar{T}\bar{A}_{ii}(t)\bar{T}^{-1})+\sum_{j\ne i}\norm{\bar{T}\bar{A}_{ij}(t)\bar{T}^{-1}}_p\le -\bar{\sigma}$,  $\forall i$ and {$\forall x\in\R^{nN}, \forall x_l \in \R^{nM}$};
    \item[C3] $\sum_j\norm{\bar{T}\bar{B}_{ij}(t)\bar{T}^{-1}}_p \le \underline{\sigma}$, $\forall i$, $\forall x\in\R^{nN}, \forall x_l \in \R^{nM}$.
\end{itemize}
where the matrices $\bar{A}_{ii}(t), \bar{A}_{ij}(t), \bar{B}_{ij}(t)$ are defined as in \Cref{equ: matrix} with state dependence omitted for brevity. {Then, we have that} the system is $\sL_\infty^p$-Input-to-State Scalable, with
\begin{align}\label{equ: scalability_bound}
\begin{split}
        &\max_i\abs{x_i(t)-x_i^\ast(t)}_p \le \frac{\kappa_p(\bar T)}{\bar{\sigma}-\underline{\sigma}}\max_i\norm{w_i(\cdot)}_{\sL_\infty^p}+ \\
        &\kappa_p(\bar T)e^{-\lambda (t-t_0)} \Bigg(\max_i\sup_{t_0-\tau_{\max}\le s \le t_0}\abs{x_i(s)-x_i^\ast(s)}_p + \\
        &\max_i\sup_{t_0-\tau_{\max}\le s \le t_0}\sum_{k=1}^m\vert r_{i,k}(s)+\\
        &\sum_{b=0}^{m-k} \frac{(m-1-b)!}{(m-k-b)!}\cdot \bar d_{i,m-1-b}\cdot s^{m-k-b}\vert_p \Bigg) , \forall N,
\end{split}
\end{align}
where $\kappa_p(\bar T):=\norm{\bar{T}}_p\norm{\bar{T}^{-1}}_p$ and $\lambda >0${, the convergence rate}, which the solution to
\begin{align}\label{equ: convergence}
    \lambda-\bar{\sigma}+\underline{\sigma}e^{\lambda\tau_{\max}}=0.
\end{align}

%    \item \textcolor{black}{$x_i(s)=\varphi_i(s), r_{i,k}(s)=\phi_{i,k}(s)$ and $x_i^\ast(s)=x_i^\ast(t_0)$, $s\in [t_0-\tau_{\max},t_0], i=1,\ldots, N, k=1,\ldots,m$.}
\end{prop}
% \begin{figure*}[b]
% \tcbset{
%         boxrule=0.5pt,
%         colframe=black
%        }

% \end{figure*}
% Proposition \Cref{prop: weighted} allows to consider situations where some of the coupling functions in \Cref{equ: control} are equal to $0$. This is useful in the special case when, for example, it is assumed that all the information to which the agents have access is delay-free. In this special case, the absence of delays implies that condition {$C3$} is satisfied with $\underline{\sigma} =0$ (when there are no delays the $\bar{B}_{ij}(t)$'s are matrices of all zeros). 

The proof of the result is given in the Appendix \ref{app:proof}. {While the proof follows the spirit of that of Proposition 1 from our previous work \cite{9353260}, the results in \cite{9353260} cannot be directly applied to the closed-loop network system \Cref{equ: dynamics} - \Cref{equ: control} due to its lack of multiplex layers, which are crucial to reject polynomial disturbances (see also Remark \ref{rem:norm}). Additionally, we note that for the application of Proposition \ref{prop: weighted} given in this paper, we fix a family of controllers and then tune the control gains so that the conditions C1-C3 are satisfied.} We now make the following considerations on the conditions.
\begin{rmk}\label{rem:norm}
From design viewpoint, if one wants to guarantee rejection of polynomial disturbances of order up to ${m-1}$, then $m$ multiplex layers need to be foreseen. That is, the control protocol \Cref{equ: control} for the $i$-th agent needs to foresee dynamics for $r_{i,k}(t)$'s with $k=1,\ldots,m$ - see the second term of the upper bound in \Cref{equ: scalability_bound}.  {Also, the factorial term is the sum of derivatives of polynomial disturbances. Essentially, with this term, we capture the fact that the polynomial disturbances vanish in time due to its multiplication with a decreasing exponential. As such, these terms are crucial to show that polynomial disturbances are rejected.} In accordance with \Cref{prop: weighted}, the protocol also guarantees that $w_i(t)$'s in \Cref{equ: disturbance} are not amplified across network.
\end{rmk}
\begin{rmk}\label{rmk: convergence}
Condition $C1$ implies that $u_i(t)=0$ at desired solution. This rather common condition (see e.g. \cite{MONTEIL2019198,9353260}) guarantees that $x^\ast(t)$ is a solution of unperturbed dynamics. This assumption is satisfied in e.g. all consensus/synchronization dynamics with diffusive-type couplings. Condition $C2$, giving an upper bound {(uniform in $t$ and $x$)} on the matrix measure of the Jacobian of delay-free part of the closed-loop network dynamics, is a diagonal dominance condition. {That is, $C2$ is a contractivity condition on the delay-free part of the dynamics (see e.g.,  \cite{centorrino2023euclidean, entrainment_2010} and references therein where contractivity is proved under a wide range of technical conditions).} Instead, condition $C3$ gives an upper bound on the norm of Jacobian of dynamics containing delays. {Finally, as we shall see, we recast the problem of fulfilling C2-C3 as an optimization problem. This problem, in order to be numerically solved, requires that the matrix measures/norms of the Jacobians in C2 and C3 are upper bounded.}

% \blue{Note that the Jacobians are state-dependent in \cref{prop: weighted} and $C2$ and $C3$ need to be satisfied $\forall x$. However, as we shall see in the optimisation setup in Appendix B, these conditions can be effectively recast as LMIs by imposing slope-restricted condition to the nonlinear couplings.}

% With $\lambda-\bar{\sigma}+\underline{\sigma}e^{\lambda\tau_{\max}}=0$, the requirement of $0\le\underline{\sigma}<\bar{\sigma}<+\infty$ in fact guarantees the existence of a positive $\lambda$ which is the convergence rate in \eqref{equ: scalability_bound}. Moreover, for a given $\tau_{\max}$, larger $\bar\sigma -\underline{\sigma}$ yields larger convergence rate and smaller upper bound induced by $w_i(t)$ in \eqref{equ: scalability_bound}.

% Moreover, the fulfilment of $C2$ and $C3$ despite their time- and state-dependence is guaranteed by setting as constraints the existence of upper bounds for the maximum possible matrix measure/norm of the Jacobian in the optimisation problem in appendix \ref{app:optimisation}.}

% \Cref{prop: weighted} implies that the matrix measure should be {\em negative enough} to compensate the presence of delays.
\end{rmk}
%Condition $C2$ is a diagonally dominance condition on the Jacobian of the delay-free part of the network dynamics.
% \begin{rmk}
% \blue{It is intrinsic in Definition \Cref{def:desired_sol} that when desired solution is achieved it must hold that $u_i(t) = 0$. This property is satisfied by any diffusive-type control protocol and is implicitly used in \cite{8370724}. Nevertheless, our technical results can be adapted to tackle situations where the desired solution is defined differently, i.e. allowing for $u_i(t)\ne 0$ at the desired solution to track time-varying references. See the design of control protocol \Cref{equ: fbl_control} for example.} 
% \end{rmk}

\begin{rmk}
Conditions $C2$ and $C3$ can be leveraged to shape the coupling functions between agents and to determine the maximum number of neighbours for each agent in the network. {Note that \Cref{prop: weighted} requires $\bar{\sigma}$ to be larger than $\underline{\sigma}$.} Moreover, if $C2$ and $C3$ are satisfied {for some $\underline{\sigma}$, $\bar{\sigma}$}, then the network is also {\em connective stable} in the sense of \cite[Chapter 2.1]{siljak2011decentralized}. Intuitively, a network is connective stable if the removal of couplings preserves stability. 
%This property and the related {\em $\gamma$-scalability} from \cite{knorn2020scalable} have been investigated for delay-free networks. 
%In \Cref{app:optimisation} we show that $C2$ and $C3$ can be fulfilled by \textcolor{magenta}{solving an optimisation problem that allows to design the control protocol for each agent independently on the other agents.}
\end{rmk}
%\GR{I would like to talk about the part in magent above.}

% \begin{remark}\label{remark: convergence}
% Following Proposition \Cref{prop: weighted}, $\lambda$ is the convergence rate of the closed-loop network. In particular, $\lambda=\bar{\sigma}$ when $\tau_{\max}=0$ (i.e. when there are no delays). From the design viewpoint, the expression in \Cref{eqn:convergence_rate} can be used to design the protocols so that the network exhibits a given, desired, convergence rate.
% \end{remark}
\begin{rmk}
\Cref{prop: weighted} generalises a number of results in the literature.  {In the special case when the network topology is a string, there are no delays and $\bar{d}_i(t)$ is constant $\forall i$, then our conditions yield these from \cite{SILVA2021109542}}.  That is, we extend the results in \cite{SILVA2021109542, MONTEIL2019198} by considering dynamic compensation of polynomial disturbances, general network topologies and delays. We also extend our prior work \cite{9353260} in which scalability is considered without  rejection of polynomial disturbances. 
%\textcolor{black}{and note that, in this special case, our conditions in Proposition \Cref{prop: weighted} become these from \cite{9353260}}.
\end{rmk}
The next result immediately follows from \Cref{prop: weighted}.

\begin{cor}\label{cor:scalability}
Consider the closed-loop network system \Cref{equ: dynamics} - \Cref{equ: control} affected by disturbances \Cref{equ: disturbance}. Assume that all the conditions in \Cref{prop: weighted} are satisfied and that, in addition, the output functions $g_i(\cdot)$ are Lipschitz. Then the system is $\sL_\infty^p$-Input-Output Scalable.
\end{cor}
\noindent\textit{Proof}: The proof, directly following from the Lipschitz hypothesis on $g_i(\cdot)$ and from \Cref{equ: scalability_bound}, is omitted here for brevity. 

%\section{Using Proposition \Cref{prop: weighted} to design multiplex protocols  for formation scalability}
\subsection{RUNNING EXAMPLE: MULTI-ROBOT FORMATION CONTROL (CONTINUE)}\label{sec: application}
% ***THE PRESENCE OF DELAYS/BROADCAST SIGNAL FROM THE LEADER SHOULD ALSO HAVE BEEN INTRODUCED EARLIER****
Following the previous part of the running example, in order to guarantee that the robotic formation satisfies R1-R4, we now design a control protocol that guarantees $\sL_\infty$-Input-to-State Scalability for \Cref{equ: dynamics_hand_position}. In order to design the protocol, we leverage \Cref{prop: weighted} and show how conditions $C1$ - $C3$ can be satisfied. We start with satisfying $C1$. Following \cite{li2013leader}, the protocol we design is of the form
\begin{align}\label{equ: fbl_control}
    \nu_i(t)=\bar{\nu}_i(t)+v_l(t),
\end{align}
where $\bar{\nu}_i(t)$ is a diffusive coupling. Following \Cref{rem:norm}, we foresee two multiplex layers in the protocol (this allows to reject the first order polynomial $\bar{d}_i(t)$ in \Cref{equ: d_bar}). Thus, we set: 
\begin{align}\label{equ: coupling_functions}
	\begin{split}
		\bar \nu_i(t)&=r_{i,1}(t)+k_{0}(\eta_l(t)-\eta_i(t)-\delta_{li}^\ast)\\
  &+k_{0}^{(\tau)}\sum_{j\in \sN_i}\psi(\eta_j(t-\tau(t))-\eta_i(t-\tau(t))-\delta_{ji}^\ast)\\
        \dot{r}_{i,1}(t)&=r_{i,2}(t)+k_{1}(\eta_l(t)-\eta_i(t)-\delta_{li}^\ast)\\
        &+k_{1}^{(\tau)}\sum_{j\in \sN_i}\psi(\eta_j(t-\tau(t))-\eta_i(t-\tau(t))-\delta_{ji}^\ast)\\
        \dot{r}_{i,2}(t)&=k_{2}(\eta_l(t)-\eta_i(t)-\delta_{li}^\ast)\\
        &+k_{2}^{(\tau)}\sum_{j\in \sN_i}\psi(\eta_j(t-\tau(t))-\eta_i(t-\tau(t))-\delta_{ji}^\ast).
	\end{split}
\end{align}
In \Cref{equ: coupling_functions}, $\psi(x):=\tanh({k_\psi} x)$ is inspired from \cite{MONTEIL2019198} and $k_{0},$ $k_{1},$ $k_{2},$ $k_{0}^{(\tau)}, k_{1}^{(\tau)}, k_{2}^{(\tau)}, {k_\psi}$ are {non-negative control gains}. Before designing the gains we note that: (i) $C1$ is satisfied by the protocol; (ii) the desired solution is a solution for the closed-loop dynamics
\begin{align}\label{equ: new_dynamics_hand_position}
	\dot{\eta}_i(t)=v_l(t)+\bar \nu_i(t)+d_i(t), \forall i.
\end{align}
Next, we design the control gains so that these fulfill $C2$ and $C3$. As shown in the Appendix \ref{app:optimisation}, the problem of finding control gains that fulfill these two conditions can be recast as the following optimisation problem:
\begin{align}\label{equ: cvx_optimisation}
	\begin{split}
		&\min_{\xi} \  \mathcal{J}\\
		&s.t. \quad k_{0}\ge0, k_{1}\ge0, k_{2}\ge0, \bar{g}_{0}\ge0, \bar{g}_{1}\ge0, \bar{g}_{2}\ge0, \\
		&\phantom{s.t. \quad} k_0+\bar g_0>0, k_1+\bar g_1>0, k_2+\bar g_2>0,\\ 
		&\phantom{s.t. \quad}\bar{\sigma}>0, \underline{\sigma}\ge 0, \bar{\sigma}-\underline{\sigma}>0, [\bar{T}\bar{A}_{ii}\bar{T}^{-1}]_s \preceq -\bar{\sigma}I_6,\\
		&\phantom{s.t. \quad}\left[\begin{matrix}\frac{\underline{\sigma}}{2\bar{N}}I_6 & (\bar{T}\tilde{B}_{ij}\bar{T}^{-1})\T \\ \bar{T}\tilde{B}_{ij}\bar{T}^{-1} & \frac{\underline{\sigma}}{2\bar{N}}I_6\end{matrix}\right]\succeq 0, \\
            &\phantom{s.t. \quad}\left[\begin{matrix}\frac{\underline{\sigma}}{2}I_6 & (\bar{T}\tilde{B}_{ii}\bar{T}^{-1})\T \\ \bar{T}\tilde{B}_{ii}\bar{T}^{-1} & \frac{\underline{\sigma}}{2}I_6\end{matrix}\right]\succeq 0.
	\end{split}
\end{align}
In the above problem, we have $\xi:=[k_{0},k_{1},k_{2},\bar g_{0},\bar g_{1},\bar g_{2}, \allowbreak \underline{\sigma}, \bar{\sigma}]$ with $\bar{g}_0={k_\psi} k_{0}^{(\tau)}, \bar{g}_1={k_\psi} k_{1}^{(\tau)}$ and $\bar{g}_2={k_\psi} k_{2}^{(\tau)}$. The matrices in the problem are defined in \Cref{equ: LMI_matrix} and \eqref{equ: blk_diag}, {where $\alpha_1, \alpha_2 \in \R$ are parameters of the coordinate transformation in \eqref{equ: blk_diag}.}
\begin{equation}\label{equ: LMI_matrix}
\begin{gathered}
    \bar{A}_{ii}=\left[\begin{matrix} -k_{0}I_2 & I_2 & 0_2\\
     			-k_{1}I_2 & 0_2 & I_2\\
     			-k_{2}I_2 & 0_2 & 0_2\end{matrix}\right],  
    \tilde{B}_{ii}=-\bar{N}\left[\begin{matrix} \bar{g}_{0} I_2 & 	0_2 & 0_2\\
     			\bar{g}_{1} I_2 & 0_2 & 0_2\\
     			\bar{g}_{2} I_2 & 0_2 & 0_2\end{matrix}\right] \\
    \tilde{B}_{ij}=\left[\begin{matrix} \bar{g}_{0} I_2& 	0_2 & 0_2\\
     			\bar{g}_{1} I_2& 0_2 & 0_2\\
     			\bar{g}_{2} I_2 & 0_2 & 0_2\end{matrix}\right],
    \bar{T}=\left[\begin{matrix}I_{2} & \alpha_{1} I_{2} &  0_2 \\
0_2 & I_{2} & \alpha_2 I_2 \\
0_2 & 0_2 & I_{2} \end{matrix}\right]
\end{gathered}
\end{equation}
In \Cref{equ: cvx_optimisation} we used the cost $\mathcal{J}:=-\bar{g}_{0}-\bar{g}_{1}-\bar{g}_{2}$, which was chosen in accordance to \cite{MONTEIL2019198} with the aim of maximizing the upper bound of the inter-robot coupling functions (other cost functions could be chosen as the steps described in the Appendix \ref{app:optimisation} are not dependent on $\mathcal{J}$). 

\begin{rmk}
For concreteness, we used \Cref{prop: weighted} for multiplex networks rejecting first order polynomials (in \Cref{sec:validation} we benchmark this with other approaches). \Cref{prop: weighted} can also be used to design networks rejecting higher order polynomials. In the Appendix \ref{SM: higher_order}, we show how the above setting is still suitable in this case. 
\end{rmk}
\begin{rmk}\label{rmk: relationship}
{Following \Cref{prop: weighted}, one can tune control parameters to tune both $\bar\sigma$ and $\underline{\sigma}$. In turn, the difference $\bar{\sigma}-\underline{\sigma}$ directly affect system's performance. In fact: (i) for a given $\tau_{\max}$, following \eqref{equ: convergence} the larger $\bar{\sigma}-\underline{\sigma}$ is, the larger $\lambda$ will be. That is, larger values of $\bar{\sigma}-\underline{\sigma}$ yield a larger convergence rate; (ii) from \eqref{equ: scalability_bound} we get that $\lim\sup_{t\rightarrow+\infty} \max_i\abs{x_i(t)-x_i^\ast(t)}_p \le \frac{\kappa_p(\bar T)}{\bar{\sigma}-\underline{\sigma}}\max_i\norm{w_i(\cdot)}_{\sL_\infty^p}$, $\forall N$. Hence, the larger $\bar{\sigma}-\underline{\sigma}$ is, the smaller the deviation induced by $w_i(t)$ at steady state will be.} 
\end{rmk}
\section{VALIDATION}\label{sec:validation}
{We now validate the protocol designed in accordance with our main theoretical results within the running example. Specifically, we present both in-silico and experimental hardware validations with real robots.} We validate protocol \Cref{equ: fbl_control} designed in accordance with our conditions, showing that it guarantees that {requirements R1-R4 are effectively fulfilled}. In the experiments, the robots need to keep a formation consisting of concentric circles (the $k$-th circle consists of $4k$ robots) and their hand positions need to move following a reference trajectory. Robots receive the velocity and position signals from the virtual leader and have access to the position information of a maximum of $\bar{N}=3$ neighbours (i.e. the closest robots). Specifically, a given robot on the $k$-th circle is connected to the robots immediately ahead and behind on the same circle and with the closest robot on circle $k-1$ (if any). The set-up we consider in our validation experiments is schematically illustrated in \Cref{fig:formation_pattern} together with the reference trajectory from the virtual leader. 
% \begin{figure}%[thbp]
% \centering
% \includegraphics[width=0.8\linewidth]{formation_pattern.eps}
% \caption{The set-up considered in the experiments of Section \Cref{sec:validation}. The reference trajectory of the hand position provided by the virtual leader is shown together with an example of  desired formation consisting of $3$ concentric circles.}
% \label{fig:formation_pattern}
% \end{figure}
In the figure, for clarity, $3$ concentric circles are shown. {We used a delay of $\tau(t)=0.1+0.1\sin t$ s} in both our simulations and hardware experiments. We first illustrate the results from the simulations and then the results obtained from the experiments on the Robotarium. {The code and data to replicate all the experiments of \Cref{sec:validation} can be found at \url{http://tinyurl.com/46xvfy7f}}.

\begin{figure}[thbp]
         \centering
         \includegraphics[width=18.5pc]{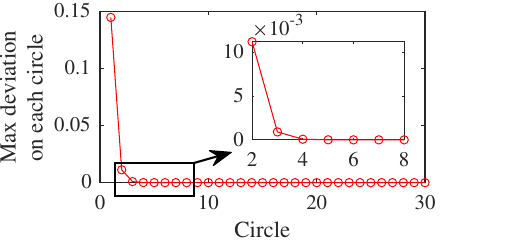}
        \caption{Maximum hand position deviation (in meters) of each circle across as number of circles increases from $1$ to $30$.}
        \label{fig: max_deviation}
\end{figure}

\begin{figure}[thbp]
\begin{subfigure}{0.5\textwidth}
\centering\includegraphics[width=18.5pc]{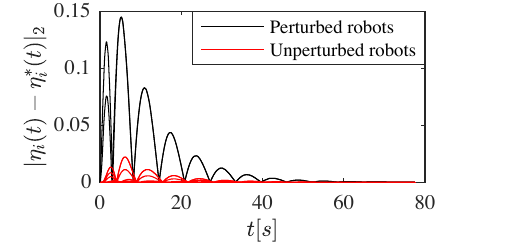}
\end{subfigure}
\begin{subfigure}{0.5\textwidth}
\centering
\includegraphics[width=18.5pc]{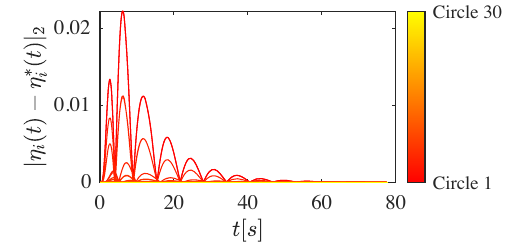}
\end{subfigure}
    \caption{Top panel: hand position deviations of all the robots (in meters). Bottom panel: hand position deviations of the unperturbed robots only (in meters). Robots on the same circle have the same color. Disturbances are the ones in \Cref{eqn:simulation_disturbances}.}
         \label{fig: all_robots}
\end{figure}

\paragraph*{{IN-SILICO VALIDATION}}
We consider a formation of $30$ circles, with two robots on circle $1$ (say, robot $1$ and robot $3$) affected by disturbances:
\begin{equation}\label{eqn:simulation_disturbances}
\begin{split}
   d_1(t)  &=\left[\begin{array}{*{20}c}
        0.04+0.4\sin(0.5t)e^{-0.1t}  \\
        0.04+0.4\sin(0.5t)e^{-0.1t}
   \end{array}\right],\\
   d_3(t)  &=\left[\begin{array}{*{20}c}
        -0.05t+0.4\sin(0.5t)e^{-0.1t}  \\
        -0.05t+0.4\sin(0.5t)e^{-0.1t}
   \end{array}\right].
\end{split}    
\end{equation}
We computed the control gains in \Cref{equ: coupling_functions} by solving the optimisation problem in \Cref{equ: cvx_optimisation} for a grid of parameters $\alpha_1$ and $\alpha_2$. We then selected the gains as the ones returning the lowest cost for each fixed pair of $\alpha$'s. By doing so, we obtained the gains $k_{0}=1.4155$, $k_{1}=1.5103$, $k_{2}=0.4803$, $k_{0}^{(\tau)}=0.642$, $k_{1}^{(\tau)}=0.872$, $k_{2}^{(\tau)}=0.425, {k_\psi}=0.1$ (corresponding to $\alpha_{1}=-0.6, \alpha_{2}=-1.6$). In \Cref{fig: max_deviation} the maximum hand position deviation is shown when the number of robots in the formation is increased, starting with a formation of $1$ circle only to a formation with $30$ circles (i.e. $1860$ robots). The figure was obtained by starting with a formation of $1$ circle and increasing at each simulation the number of circles. We recorded at each simulation the maximum hand position deviation for each robot on a given circle and finally plotted the largest deviation on each circle across all the simulations. The figure clearly shows that the polynomial components of the disturbances in \Cref{eqn:simulation_disturbances} are rejected by the multiplex control protocol and the residual disturbances are not amplified through the formation. We also report the behaviour of the full formation with $30$ circles in  \Cref{fig: all_robots}. 
\begin{figure}[thbp]
         \centering
         \includegraphics[width=0.95\linewidth]{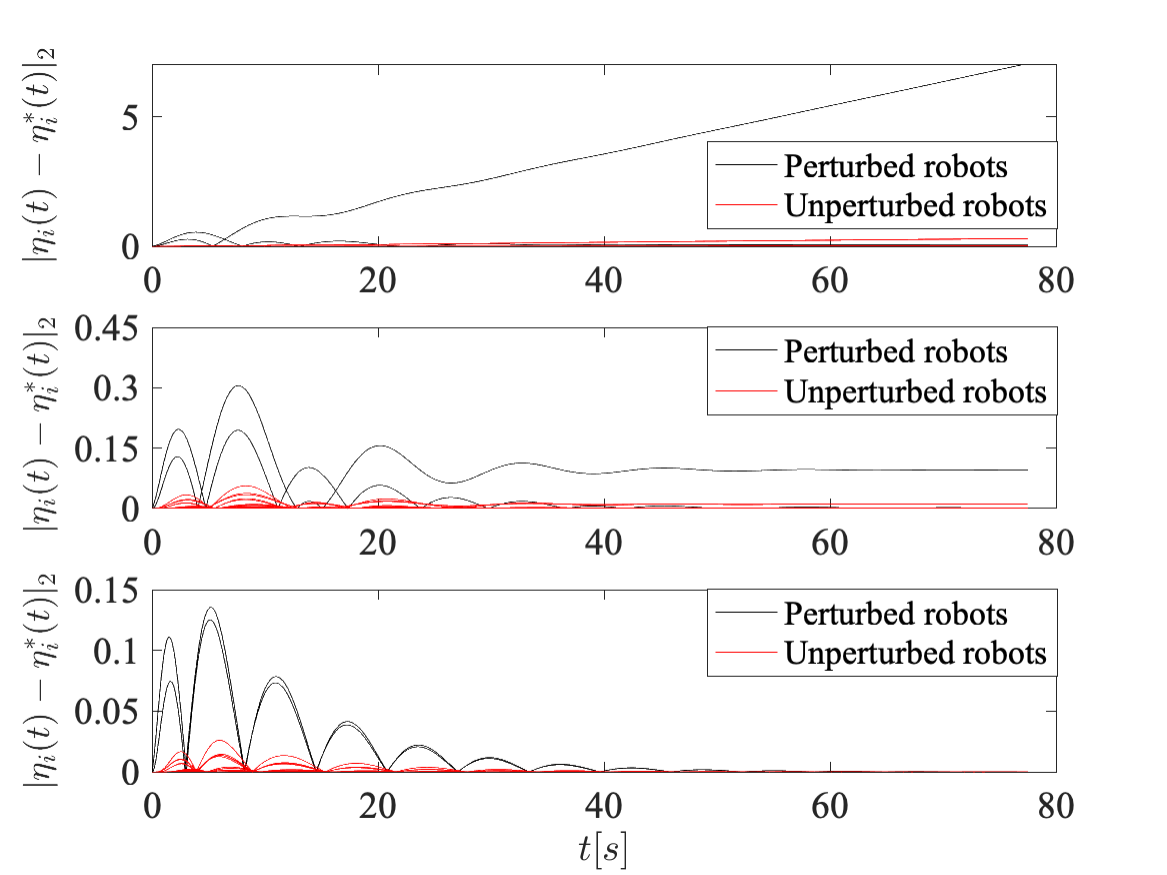}
        \caption{\black{Hand position deviation when 
        the control protocol is designed according to \cite{9353260} (top), \cite{SILVA2021109542} (middle) and when the protocol in \Cref{equ: fbl_control} is used (bottom).}}
        \label{fig: comparison}
\end{figure}
Both panels of the figure confirm that, in accordance with our theoretical results, the protocol allows the robots to keep the desired formation and track the reference trajectory, while rejecting the polynomial components of the disturbances and prohibiting the amplification of the residual disturbances.

\begin{figure}
         \centering
         \includegraphics[width=18.5pc]{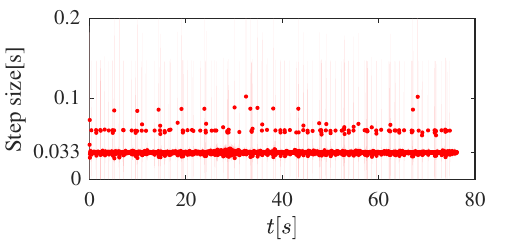}
        \caption{Measurement of the step size on the Robotarium hardware. Solid markers represent the average step size from $10$ sets of experiments while the shaded area represents the confidence interval corresponding to the standard deviation.}
        \label{fig: stepsize}
\end{figure}

Finally, before presenting the validation results on Robotarium, we benchmark the performance of our protocol \Cref{equ: fbl_control} to control \Cref{equ: dynamics_hand_position} with those obtained following \cite{9353260} and \cite{SILVA2021109542}. {We pick \cite{9353260} and \cite{SILVA2021109542} as these are the two works that are most related to our results.} Since the design conditions from \cite{SILVA2021109542} are tailored towards networks with a (bi-directional) string topology, we compared performance of the protocols using such a topology. To this aim we again considered the formation control problem for the formation of \Cref{fig:formation_pattern} with $30$ circles this time with each robot bidirectionally coupled to the robots before and after (that is, robot $i$ in the formation was coupled to robot $i-1$ and $i+1$, if any). The disturbance considered in the simulation was again the one in \Cref{eqn:simulation_disturbances} and the network had no delays (the results from \cite{SILVA2021109542} apply to delay-free networks). The time evolution of hand position deviations for robots controlled by our control protocol is shown in the bottom panel of \Cref{fig: comparison}. The simulation results, consistently with our theoretical findings, show that the desired scalability property is achieved. The middle panel of the figure instead shows the time evolution of the hand position deviations when a control protocol designed according to \cite{SILVA2021109542} is used. In this case, as shown in the panel, the protocol is not able to reject the polynomial disturbances. Finally, we also benchmarked the performance of our protocol with a protocol designed following \cite{9353260}. To do so, we considered a situation where the network is affected by time-varying delay $\tau(t)=0.1+0.1\sin t$ s. In the top panel of \Cref{fig: comparison} the time evolution of the deviation of the hand position is shown when a protocol designed according to \cite{9353260} is used. The panel clearly shows that the first order polynomial disturbance is not rejected by this protocol.
%, in the top panel of such a figure e report in Figure \Cref{fig: all_robots} the hand position deviation of all robots. Clearly the polynomial components in the disturbances $d_1(t), d_2(t)$ are fully rejected as no constant nor increasing position deviation are witnessed for perturbed robots. Figure \Cref{fig: unperturbed_robots} illustrates the hand position deviation of the unperturbed robots only, with robots on the same circle have the same colour. It shows that the hand position deviation of circle $30$ is negligible compared to circle $1$ where external disturbances enter the network. This conforms to our theoretical prediction that scalability property prohibits the amplification of the perturbation caused by the residual components in $d_1(t), d_2(t)$.
\paragraph*{EXPERIMENTAL VALIDATION}
We further validate our results by carrying out experiments on Robotarium, which provides both hardware infrastructure and a high-fidelity simulator of the hardware. In the experiments, a formation of $2$ concentric circles (hence with $12$ robots) is considered and, for consistency with our previous set of simulations, $2$ robots on circle $1$ are perturbed by the disturbances given in \Cref{eqn:simulation_disturbances}. The Robotarium documentation\footnote{\url{https://tinyurl.com/3rajpnep}} reports a nominal step size (for both the simulator and the hardware infrastructure) of $0.033$s. Since the step size is used to implement the multiplex layers in \Cref{equ: coupling_functions} as a first step we measured the actual step size in the hardware infrastructure. The result is given in \Cref{fig: stepsize}. 
\begin{figure}[thbp]
\begin{subfigure}{0.5\textwidth}
    \centering
         \includegraphics[width=18.5pc]{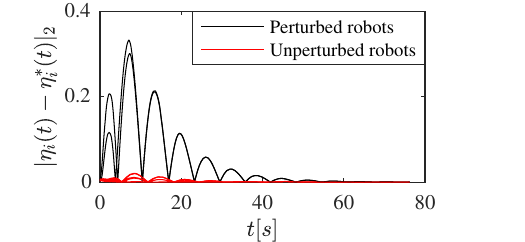} 
\end{subfigure}
\begin{subfigure}{0.5\textwidth}
    \centering
    \includegraphics[width=18.5pc]{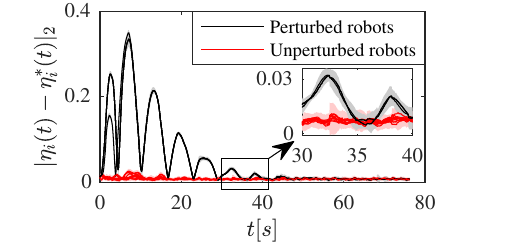}
\end{subfigure}        
         \caption{Top panel: hand position deviations of the robots (in meters) from Robotarium simulator. Bottom panel: hand position deviations (in meters) from the hardware experiments. Solid lines represent the average hand position deviations across $10$ sets of experiments; the shaded area represents the confidence interval corresponding to the standard deviation \black{(part of the plot, between $30$s and $40$s, has been magnified to enhance visibility )}. See \url{http://tinyurl.com/46xvfy7f} for the {\em animated} version of the plot.}
     \label{fig: hand_position_deviation}
\end{figure}
Such a figure reports the average step size we measured using built-in timing functions across $10$ experiments\footnote{See our code at \url{http://tinyurl.com/46xvfy7f} for the details on these measurements.}. In the same figure, the shaded area represents the confidence interval corresponding to the standard deviation. As illustrated in the figure, while the average step size is indeed around $0.033$s and consistent with the nominal value, it also introduces some variability in the experiments. Such variability leads to the observation of a gap between the results obtained from the Robotarium simulator and the actual experimental results. This phenomenon is essentially due to the fact that in the Robotarium experiments we could only use the nominal step-size of $0.033$s (and not the actual step size) for the implementation of the dynamics of the multiplex layers in the protocol. We decided to mitigate this {\em simulation-to-reality} gap by reducing the gains for the couplings of the multiplex layers. Hence, in the experimental results presented next, we impose that the control gains of the multiplex layers (i.e., layer $1$ and layer $2$) are smaller than the gains of layer $0$. This was done by solving the optimisation problem in \Cref{equ: cvx_optimisation} this time with the following additional constraints: $k_0\ge2k_1, k_0\ge 2k_2, \bar g_0\ge2\bar g_1, \bar g_0\ge 2\bar g_2$. As an outcome of this process, we obtained the following gains: $k_{0}=1.2674$, $k_{1}=0.6312$, $k_{2}=0.133$, $k_{0}^{(\tau)}=0.325$, $k_{1}^{(\tau)}=0.162$, $k_{2}^{(\tau)}=0.06, {k_\psi}=0.1$ (which correspond to $\alpha_{1}=-1.1, \alpha_{2}=-2.6$). We then validated the control protocol with this choice of parameters by first leveraging the Robotarium simulator and the results, consistent with our theoretical findings, are shown in the top panel of \Cref{fig: hand_position_deviation}. Next, we validated the control protocol on the Robotarium hardware infrastructure and the outcome from these experiments are shown in the bottom panel of \Cref{fig: hand_position_deviation}. In the figure, which was obtained from a set of $10$ experiments, the solid lines are the robots' average hand position deviations and the shaded area represents the confidence interval corresponding to the standard deviation. The behaviour of the hardware experiments is in agreement with the one obtained from the simulator. Both panels show that, in accordance with \Cref{prop: weighted}, our multiplex control protocol allowed the robots to keep the desired formation and track the reference trajectory, while rejecting the polynomial components of the disturbances and ensuring {non-amplification of the residual disturbances}. A video recording of the experiment is available at \url{http://tinyurl.com/46xvfy7f}.

\section{CONCLUSIONS AND FUTURE WORK}\label{sec:conclusions}
We considered nonlinear network systems affected by delays and, for these systems, we presented a set of sufficient conditions to guarantee a scalability property. This property implies that the network achieves some desired behaviour, while simultaneously guaranteeing rejection of polynomial disturbances and the non-amplification of other piece-wise continuous disturbances across the nodes. The conditions, which yielded a multiplex network architecture, were then used to design protocols for formation control. The effectiveness of the protocols was then illustrated via both in-silico and hardware validations. With our future work, we are  interested in investigating scalability of networks with heterogeneous delays. {In this setting, the error dynamics cannot be written in the form used to prove Proposition \ref{prop: weighted} and the key challenge is then that of proving contractivity (using the structured norm considered in this paper) for this different dynamics.} We also plan to study network systems evolving over arbitrary time domains. These {\em time scales} dynamics \cite{1531-3492_2021188} can be used to model discrete-time systems with non-uniform sampling times. Hence, the time scales formalism might be useful to tackle situations, also observed in \Cref{sec:validation}, where the step size is non-uniform and not known a-priori. {On a related note, we are also interested in extending our scalability results to consider sampled-data controllers and stochastic disturbances  \cite{chen2021mean, chen2023controller, yao2021stochastic}. Finally, also in view of improving the numerical tools made available on our GitHub, we are also interested in investigating, in general settings, the computational complexity of the optimisation problems required to fulfil C2-C3.}

\section*{APPENDIX}
\subsection{PROOF OF PROPOSITION \ref{prop: weighted}}\label{app:proof}
We start with augmenting the state of the original dynamics by defining $z_i(t):=[x_i\T(t), \zeta_{i,1}\T(t), \zeta_{i,2}\T(t),\cdots,\zeta_{i,m}\T(t)]\T$ where
\begin{align*}
    \begin{split}
        \zeta_{i,k}(t)&=r_{i,k}(t) + \sum_{b=0}^{m-k} \frac{(m-1-b)!}{(m-k-b)!}\cdot \bar d_{i,m-1-b}\cdot t^{m-k-b},  
    \end{split}
\end{align*}
$k=1,\cdots,m$. In these new coordinates the dynamics of the network system becomes
%\GR{Need to double check together the expressions below}
\begin{align}\label{equ: augmented_dynamics}
    \dot{z}_i(t)=\tilde{f}_i(z_i,t)+\tilde v_i(z,t)+ \tilde w_i(t),
\end{align}
where $\tilde f_i(z_i,t) =[f_i\T(x_i,t),0_{1\times n},\ldots,0_{1\times n}]\T$, $\tilde w_i(t)=[w_i\T(t),0_{1\times n},\ldots,0_{1\times n}]\T$, $\tilde v_i(z,t)\allowbreak=v_i(z,t)+v_i^{(\tau)}(z,t)$ with
{\begin{align*}
    \begin{split}
        v_i(z,t) = \left[\begin{array}{*{20}c}
h_{i,0}(x_i,\{x_j\}_{j\in\sN_i},x_l,t) + \zeta_{i,1}(t) \\
h_{i,1}(x_i,\{x_j\}_{j\in\sN_i},x_l,t) + \zeta_{i,2}(t)\\
\vdots \\
h_{i,m-1}(x_i,\{x_j\}_{j\in\sN_i},x_l,t) + \zeta_{i,m}(t)\\
h_{i,m}(x_i,\{x_j\}_{j\in\sN_i},x_l,t) \end{array}\right], 
         \end{split}
\end{align*}  
and
\begin{align*}
    \begin{split}
        v_i^{(\tau)}(z,t) = \left[\begin{array}{*{20}c}
h_{i,0}^{(\tau)}(x_i,\{x_j\}_{j\in\sN_i},x_l,t)\\
h_{i,1}^{(\tau)}(x_i,\{x_j\}_{j\in\sN_i},x_l,t)\\
\vdots \\
h_{i,m}^{(\tau)}(x_i,\{x_j\}_{j\in\sN_i},x_l,t)\end{array}\right].
    \end{split}
\end{align*}}
Condition $C1$ implies that $x^\ast(t)$ is a solution of the unperturbed dynamics, i.e. $x^\ast(t)$ is a solution of \Cref{equ: dynamics} when there are no disturbances. Moreover, when there are no disturbances, in the augmented dynamics, the solution $z_i^\ast(t):=[x_i^{\ast\mathsf{T}}(t), 0_{1\times n}, \ldots, 0_{1\times n}]\T$ satisfies $\dot{z}^\ast_i(t)=\tilde f_i(z_i^\ast,t)$ with $\tilde f_i(z_i^\ast,t):= \allowbreak [f_i\T(x_i^\ast,t),0_{1\times n}, \ldots, 0_{1\times n}]\T$. Hence, the dynamics of state deviation (i.e. the error) $e_i(t)=z_i(t)-z_i^\ast(t)$ is given by
        $\dot{e}_i(t)=\tilde f_i(z_i,t)-\tilde f_i(z_i^\ast,t)+\tilde v_i(z,t)+\tilde w_i(t).$
Following \cite{1083507}, we let $\eta_i(\rho)=\rho z_i+(1-\rho)z_i^\ast$, $\eta(\rho)=[\eta_1\T(\rho),\ldots,\eta_N\T(\rho)]\T$ and then rewrite the error dynamics as
    $\dot{e}(t)=A(t)e(t)+B(t)e(t-\tau(t))+\tilde w(t),$
where $\tilde w(t)=[\tilde w_1\T(t),\ldots,\tilde w_N\T(t)]\T$ and $A(t)$ has entries: (i) $A_{ij}(t)=\int_0^1 J_{v_i}(\eta_j(\rho),t)d\rho$; (ii) $A_{ii}(t)=\int_0^1 \left(J_{\tilde f_i}(\eta_i(\rho),t)+J_{v_i}(\eta_i(\rho),t)\right)d\rho$. Similarly, $B(t)$ has entries: $B_{ij}(t)=\int_0^1 J_{v_i^{(\tau)}}(\eta_j(\rho),t)d\rho$. In the above expressions, the Jacobian matrices are defined as $J_{\tilde f_i}(\eta_i,t) := \frac{\partial \tilde f_i(\eta_i,t)}{\partial \eta_i}$, $J_{v_i}(\eta_i,t) := \frac{\partial v_i(\eta,t)}{\partial \eta_i}$, $J_{v_i^{(\tau)}}(\eta_i,t) :=\frac{\partial v_i^{(\tau)}(\eta,t)}{\partial \eta_i}$. Now, consider the coordinate transformation $\tilde{z}(t) :=Tz(t)$ and  $\tilde{e}(t):=Te(t)$, we have
\begin{align}\label{equ: transformed_error}
    \dot{\tilde{e}}(t)=TA(t)T^{-1}\tilde{e}(t)+TB(t)T^{-1}\tilde{e}(t-\tau(t))+T\tilde w(t).
\end{align}
Let $\abs{x}_G := \left\vert\abs{x_1}_p,\ldots,\abs{x_N}_p \right\vert_{\infty}$. Then, by taking the Dini derivative of $\abs{\tilde{e}(t)}_{G}$ we may continue as follows
\begin{align*}
    \begin{split}
        &D^+\abs{\tilde{e}(t)}_{G}=\limsup_{h\rightarrow 0^+} \frac{1}{h}\left(\abs{\tilde{e}(t+h)}_{G}-\abs{\tilde{e}(t)}_{G}\right)\\
        =&\limsup_{h\rightarrow 0^+} \frac{1}{h}\bigl(\abs{\tilde{e}(t)+hTA(t)T^{-1}\tilde{e}(t)+\\
        &hTB(t)T^{-1}\tilde{e}(t-\tau(t))+hT\tilde w(t)}_{G}-\abs{\tilde{e}(t)}_{G}\bigr)\\
        \le& \limsup_{h\rightarrow 0^+} \frac{1}{h}\left(\norm{I+hTA(t)T^{-1}}_{G}-1\right)\abs{\tilde{e}(t)}_{G}+\\
        &\norm{TB(t)T^{-1}}_{G}\abs{\tilde{e}(t-\tau(t))}_{G}+\abs{T\tilde w(t)}_{G}\\
        % &\le \mu_{G}(TA(t)T^{-1})\abs{\tilde{e}(t)}_{G} + \norm{TB(t)T^{-1}}_{G} \sup_{t-\tau_{\max}\le s \le t}\abs{\tilde{e}(s)}_{G}+\abs{T\tilde w(t)}_{G}\\
        \le& \mu_{G}(TA(t)T^{-1})\abs{\tilde{e}(t)}_{G}+ \\
        & \norm{TB(t)T^{-1}}_{G} \sup_{t-\tau_{\max}\le s \le t}\abs{\tilde e(s)}_{G}+\norm{T}_G\max_i \norm{\tilde w_i(\cdot)}_{\sL_\infty^p}.
    \end{split}
\end{align*}
Next, we find upper bounds for $\mu_{G}(TA(t)T^{-1})$ and $\norm{TB(t)T^{-1}}_{G}$ which allow us to apply \Cref{prop:halanay}. First, we give the expression of the matrix $\bar{A}(t)$ which have entries: (i) $\bar{A}_{ii}(t)=J_{\tilde f_i}(z_i,t)+J_{v_i}(z_i,t)$; (ii) $\bar{A}_{ij}(t)=J_{v_i}(z_j,t)$, and $\bar{B}(t)$ has entries: $\bar{B}_{ij}(t)=J_{v_i^{(\tau)}}(z_j,t)$. Then, by sub-additivity of matrix measures and matrix norms, we get
$\mu_{G}(TA(t)T^{-1}) \le \int_0^1 \mu_{G}(T \bar A(t)T^{-1})d\rho$ and $\norm{TB(t)T^{-1}}_{G} \le \int_{0}^1\norm{T\bar B(t)T^{-1}}_{G}d\rho$ (see also Lemma $3.4$ in \cite{1531-3492_2021188}). Moreover, from \Cref{lem:matrix norm} it follows that
\begin{itemize}
    \item $\mu_{G}(T\bar A(t)T^{-1}) \le \max_i\Bigl\{\mu_p(\bar{T} \bar A_{ii}(t)\bar{T}^{-1})+\sum_{j\ne i}\norm{\bar{T} \bar A_{ij}(t)\bar{T}^{-1}}_p\Bigr\},$
    \item $\norm{T\bar B(t)T^{-1}}_{G} \le \max_i\Bigl\{\sum_j\norm{\bar{T} \bar B_{ij}(t)\bar{T}^{-1}}_p\Bigr\}.$
\end{itemize}
{Note now that, since} conditions $C2$ and $C3$ {are satisfied by hypotheses, we have that, for all $i$:
\begin{itemize}
    \item $\mu_p(\bar T\bar A_{ii}(t)\bar T^{-1})+\sum_{j\ne i}\norm{\bar T \bar A_{ij}(t)\bar T^{-1}}_p \le -\bar{\sigma},$
    \item $\sum_j\norm{\bar T \bar B_{ij}(t)\bar T^{-1}}_p \le \underline{\sigma},$
\end{itemize}}
\noindent for some $0\le\underline{\sigma}<\bar{\sigma}<+\infty$. {Hence, it follows that}
\begin{itemize}
    \item $\max_i\Bigl\{\mu_p(\bar T\bar A_{ii}(t)\bar T^{-1})+\sum_{j\ne i}\norm{\bar T \bar A_{ij}(t)\bar T^{-1}}_p\Bigr\} \le -\bar{\sigma},$
    \item $\max_i\Bigl\{\sum_j\norm{\bar T \bar B_{ij}(t)\bar T^{-1}}_p\Bigr\} \le \underline{\sigma},$
\end{itemize}
This implies that
\begin{align}
  \mu_G(TA(t)T^{-1})+\norm{TB(t)T^{-1}}_G\le -\bar{\sigma} + \underline{\sigma}:= -\sigma,  
\end{align}
and \Cref{prop:halanay} then yields 
\begin{align*}
    \abs{\tilde{e}(t)}_{G}&\le \sup_{t_0-\tau_{\max}\le s \le t_0}\abs{\tilde{e}(s)}_{G} e^{-\lambda (t-t_0)} \\
    &+ \frac{\norm{T}_G}{\bar{\sigma}-\underline{\sigma}}\max_i \norm{\tilde w_i(\cdot)}_{\sL_\infty^p},
\end{align*}
with $\lambda$ defined as in the statement of the proposition. Since $\tilde{e}(t)=Te(t)$ we get $\abs{e(t)}_G \le \norm{T^{-1}}_G\abs{\tilde{e}(t)}_G$ and $\abs{\tilde{e}(t)}_G \le \norm{T}_G\abs{e(t)}_G$. We also notice that the definition of $\tilde w_i(t)$ implies that $\norm{\tilde w_i(\cdot)}_{\sL_\infty^p}=\norm{w_i(\cdot)}_{\sL_\infty^p}$. Hence
\begin{align*}
    \abs{e(t)}_G&\le  \norm{T^{-1}}_G\norm{T}_G\biggl(\sup_{t_0-\tau_{\max}\le s \le t_0}\abs{e(s)}_{G} e^{-\hat\lambda (t-t_0)}\\
    &+ \frac{1}{\bar{\sigma}-\underline{\sigma}}\max_i \norm{w_i(\cdot)}_{\sL_\infty^p}\biggr).
\end{align*}
\Cref{lem:matrix norm} yields 
\begin{align}
\norm{T}_G\norm{T^{-1}}_G\le\norm{\bar{T}}_p\norm{\bar{T}^{-1}}_p:=\kappa_p(\bar T)    
\end{align}
and we note that $\abs{e_i(t)}_p=\vert [x_i\T(t)-x_i^{\ast\mathsf{T}}(t), \zeta_{i,1}\T(t), \ldots, \allowbreak\zeta_{i,m}\T(t) {]}\vert_p \ge \vert[ x_i\T(t)-x_i^{\ast\mathsf{T}}(t), 0_{1\times n}, \ldots, 0_{1\times n}]\vert_p=\abs{x_i(t)-x_i^\ast(t)}_p,$
% \begin{align*}
%     \abs{e_i(t)}_p&=\vert [x_i\T(t)-x_i^{\ast\mathsf{T}}(t), \zeta_{i,1}\T(t), \ldots, \allowbreak\zeta_{i,m}\T(t) \vert_p \\
%     &\ge \vert[ x_i\T(t)-x_i^{\ast\mathsf{T}}(t), 0_{1\times n}, \ldots, 0_{1\times n}]\vert_p=\abs{x_i(t)-x_i^\ast(t)}_p,
% \end{align*}
% $$\abs{e_i(t)}_p=\vert [x_i\T(t)-x_i^{\ast\mathsf{T}}(t), \zeta_{i,1}\T(t), \ldots, \allowbreak\zeta_{i,m}\T(t) \vert_p \ge \vert[ x_i\T(t)-x_i^{\ast\mathsf{T}}(t), 0_{1\times n}, \ldots, 0_{1\times n}]\vert_p=\abs{x_i(t)-x_i^\ast(t)}_p,$$ 
and $\abs{e_i(t_0)}_p=\vert[x_i\T(t_0)-x_i^{\ast\mathsf{T}}(t_0), \zeta_{i,1}\T(t_0), \ldots, \allowbreak \zeta_{i,m}\T(t_0)]\vert_p \le \vert x_i(t_0)-x_i^\ast(t_0)\vert_p + \sum_{k=1}^m \abs{\zeta_{i,k}(t_0)}_p.$
% \begin{align*}
%     \abs{e_i(t_0)}_p&=\vert[x_i\T(t_0)-x_i^{\ast\mathsf{T}}(t_0), \zeta_{i,1}\T(t_0), \ldots, \allowbreak \zeta_{i,m}\T(t_0)]\vert_p \\
%     &\le \vert x_i(t_0)-x_i^\ast(t_0)\vert_p + \sum_{k=1}^m \abs{\zeta_{i,k}(t_0)}_p.
% \end{align*}
% $$\abs{e_i(t_0)}_p=\vert[x_i\T(t_0)-x_i^{\ast\mathsf{T}}(t_0), \zeta_{i,1}\T(t_0), \ldots, \allowbreak \zeta_{i,m}\T(t_0)]\vert_p \le \vert x_i(t_0)-x_i^\ast(t_0)\vert_p +\vert\zeta_{i,1}(t_0)\vert_p +\cdots + \vert\zeta_{i,m}(t_0)\vert_p.$$ 
% Hence, $\abs{x_i(t)-x_i^\ast(t)}_p \le \abs{e_i(t)}_p$ and $\abs{e_i(t_0)}_p \le \abs{x_i(t_0)-x_i^\ast(t_0)}_p+\sum_{k=1}^m \abs{\zeta_{i,k}(t_0)}_p$. 
We then finally obtain the upper bound of the state deviation
\begin{align}
\begin{split}
    &\max_i\abs{x_i(t)-x_i^\ast(t)}_p \le \\
    &{\kappa_p(\bar T)e^{-\hat\lambda (t-t_0)}\Bigg(} \max_i\sup_{t_0-\tau_{\max}\le s \le t_0}\abs{x_i(s)-x_i^\ast(s)}_p\\   
    &+\max_i\sup_{t_0-\tau_{\max}\le s \le t_0}\sum_{k=1}^m\vert r_{i,k}(s)\\
    &+\sum_{b=0}^{m-k} \frac{(m-1-b)!}{(m-k-b)!}\cdot \bar d_{i,m-1-b}\cdot s^{m-k-b}\vert_p\Bigg)\\
    &+\frac{\kappa_p(\bar T)}{\bar{\sigma}-\underline{\sigma}}\max_i\norm{w_i(\cdot)}_{\sL_\infty^p}, \forall N.
\end{split}
\end{align}

\subsection{FULFILLING C2 AND C3 VIA OPTIMISATION}\label{app:optimisation} 
{We choose $\abs{x}_G := \abs{\left[ \abs{x_1}_2,\ldots,\abs{x_N}_{2}\right]}_\infty$.} The optimisation problem in \Cref{equ: cvx_optimisation} was obtained by noticing that $C2$ and $C3$ can be fulfilled by solving the following optimisation problem:
\begin{align}\label{equ: optimisation}
	\begin{split}
		&\min_{\bar{\xi}} \ \mathcal{J}\\
		%(k_{0},k_{1},k_{2},k_{0}^{(\tau)},k_{1}^{(\tau)},k_{2}^{(\tau)}, k^\psi)\\
		&s.t. \quad k_{0}\ge0, k_{1}\ge 0, k_{2}\ge 0, k_{0}^{(\tau)}\ge 0, k_{1}^{(\tau)}\ge 0, k_{2}^{(\tau)}\ge 0, \\
		&\phantom{s.t. \quad} {k_\psi}>0, k_0+k_0^{(\tau)}>0, k_1+k_1^{(\tau)}>0, k_2+k_2^{(\tau)}>0, \\
		&\phantom{s.t. \quad} \bar{\sigma}>0, \underline{\sigma}\ge 0, \bar{\sigma}-\underline{\sigma}>0, \mu_2(\bar{T}\bar{A}_{ii}\bar{T}^{-1})\le -\bar{\sigma},\\
		&\phantom{s.t. \quad}  \sum_{j\in\sN_i}\norm{\bar{T}\bar{B}_{ij}(t)\bar{T}^{-1}}_2+\norm{\bar{T}\bar{B}_{ii}(t)\bar{T}^{-1}}_2\le \underline{\sigma},
	\end{split}
\end{align}
where the decision variables are $\bar{\xi}:=[k_{0},k_{1},k_{2},k_{0}^{(\tau)},k_{1}^{(\tau)}, \allowbreak k_{2}^{(\tau)}, {k_\psi},  \underline{\sigma}, \bar{\sigma}, \alpha_{1}, \alpha_{2}]$ and the cost is defined as in Section IV-\ref{sec: application}. The matrices $\bar{B}_{ii}, \bar{B}_{ij}$ are given in \Cref{equ: robot_matrix} and $\bar{A}_{ii}$ is given in \Cref{equ: LMI_matrix}, in accordance with \Cref{prop: weighted} while the transformation matrix $\bar{T}$ is also given in \Cref{equ: LMI_matrix}.
\begin{equation}\label{equ: robot_matrix}
\begin{split}
    \bar{B}_{ii}(t)&=\abs{\sN_i}\left[\begin{matrix} \bar{g}_0 I_2  & 	0_2 & 0_2\\
     			\bar{g}_1 I_2  & 0_2 & 0_2\\
     			\bar{g}_2 I_2  & 0_2 & 0_2\end{matrix}\right]\frac{\partial \tanh(\eta_j-\eta_i-\delta_{ji}^\ast)}{\partial \eta_i}\\
    \bar{B}_{ij}(t)&=\left[\begin{matrix} \bar{g}_0 I_2 & 	0_2 & 0_2\\
     			\bar{g}_1 I_2 & 0_2 & 0_2\\
     			\bar{g}_2 I_2 & 0_2 & 0_2\end{matrix}\right]\frac{\partial \tanh(\eta_j-\eta_i-\delta_{ji}^\ast)}{\partial \eta_j}
\end{split} 
\end{equation}
In order to find the control gains, we propose to solve the optimisation problem for fixed $\alpha_{1}, \alpha_{2}$. Further, in order to obtain a suitable formulation for the optimisation, we recast the constraints in \Cref{equ: optimisation} as LMIs as follows. First, by definition, $\mu_2(\bar{T}\bar{A}_{ii}\bar{T}^{-1})\le -\bar{\sigma}$ is equivalent to $[\bar{T}\bar{A}_{ii}\bar{T}^{-1}]_s \preceq -\bar{\sigma}I_6$. Moreover, the constraint  $\sum_{j\in\sN_i}\norm{\bar{T}\bar{B}_{ij}(t)\bar{T}^{-1}}_2+\norm{\bar{T}\bar{B}_{ii}(t)\bar{T}^{-1}}_2\le \underline{\sigma}$ is satisfied if we impose that  $\norm{\bar{T}\bar{B}_{ii}(t)\bar{T}^{-1}}_2 \le \frac{\underline{\sigma}}{2}$ and, simultaneously, $\norm{\bar{T}\bar{B}_{ij}(t)\bar{T}^{-1}}_2 \le \frac{\underline{\sigma}}{2\bar{N}}, \forall j\in \sN_i$. In turn, since $-1\le \frac{\partial \tanh(\eta_j-\eta_i-\delta_{ji}^\ast)}{\partial \eta_i}\le 0$ and $\abs{\sN_i} \le \bar{N}$ we have (by means of the absolutely homogeneous property for matrix norms) that $\norm{\bar{T}\bar{B}_{ii}(t)\bar{T}^{-1}}_2 \le \norm{\bar{T}\tilde{B}_{ii}\bar{T}^{-1}}_2$ with $\tilde{B}_{ii}$ defined in \Cref{equ: LMI_matrix}. Analogously, we have that $\norm{\bar{T}\bar{B}_{ij}(t)\bar{T}^{-1}}_2 \le \norm{\bar{T}\tilde{B}_{ij}\bar{T}^{-1}}_2$, with $\tilde{B}_{ij}$ defined in \Cref{equ: LMI_matrix}. Hence, the constraints on the norm in \Cref{equ: optimisation} are satisfied if $\norm{\bar{T}\tilde{B}_{ii}\bar{T}^{-1}}_2 \le \frac{\underline{\sigma}}{2}$ and $\norm{\bar{T}\tilde{B}_{ij}\bar{T}^{-1}}_2 \le \frac{\underline{\sigma}}{2\bar{N}}, \forall j\in \sN_i$ which, following \cite[Example 4.6.3]{boyd2004convex}, can be written as $$\left(\frac{\underline{\sigma}}{2\bar{N}}\right)^2 I_6 - (\bar{T}\tilde{B}_{ij}\bar{T}^{-1})\T (\bar{T}\tilde{B}_{ij}\bar{T}^{-1}) \succeq 0$$ $$\left(\frac{\underline{\sigma}}{2}\right)^2 I_6 - (\bar{T}\tilde{B}_{ii}\bar{T}^{-1})\T (\bar{T}\tilde{B}_{ii}\bar{T}^{-1}) \succeq 0$$ Now, by means of Schur complement, this pair of inequalities is equivalent to 
$$\left[\begin{matrix}\frac{\underline{\sigma}}{2\bar{N}}I_6 & (\bar{T}\tilde{B}_{ij}\bar{T}^{-1})\T \\ \bar{T}\tilde{B}_{ij}\bar{T}^{-1} & \frac{\underline{\sigma}}{2\bar{N}}I_6\end{matrix}\right]\succeq 0,$$
$$\left[\begin{matrix}\frac{\underline{\sigma}}{2}I_6 & (\bar{T}\tilde{B}_{ii}\bar{T}^{-1})\T \\ \bar{T}\tilde{B}_{ii}\bar{T}^{-1} & \frac{\underline{\sigma}}{2}I_6\end{matrix}\right]\succeq 0.$$
Since the cost is {linear} and the constraints are LMIs, the optimisation problem in \Cref{equ: cvx_optimisation} is convex for fixed $\alpha$'s. {This is exploited in \Cref{sec:validation}.}
%\SX{I agree with this change.}

\subsection{REJECTING HIGHER ORDER POLYNOMIALS}\label{SM: higher_order}
%\SX{added in red statement on transformation matrix and the time-invariance of matrices.}
% \begin{figure*}[b!]
% \tcbset{boxrule=0.5pt,
%         colframe=black}
% \end{figure*}
The optimisation setting of Section IV-\ref{sec: application} is also applicable to the design progress of multiplex control protocols able to reject polynomials of arbitrary (say, $m-1$) order. We recall that, to reject such a disturbance the protocol needs to have $m$ multiplex layers (\Cref{rem:norm}). Hence, we design the protocol \Cref{equ: control} but with $m$ layers. The coupling functions for the $n$-th layer, $n\in\{0,\ldots,m\}$, are, in analogy to \Cref{equ: coupling_functions}:
\begin{align}
    \begin{split}
        h_{i,n}(\eta_i,\{\eta_j\}_{j\in\sN_i},\eta_l,t)&=k_{n}(\eta_l-\eta_i-\delta_{li}^\ast),\\ 
		h_{i,n}^{(\tau)}(\eta_i,\{\eta_j\}_{j\in\sN_i},\eta_l,t)&=k_{n}^{(\tau)}\sum_{j\in \sN_i}\psi(\eta_j-\eta_i-\delta_{ji}^\ast)
    \end{split}
\end{align}
where $\psi(x):=\tanh({k_\psi}x)$ as in \Cref{equ: coupling_functions} and $k_{n}, k_{n}^{(\tau)}, {k_\psi}$ are the control parameters to be designed.  Now, $C2$ and $C3$ can be fulfilled by solving the following  problem: 
\begin{align}\label{equ: optimisation_general}
	\begin{split}
		&\min_{\bar{\xi}} \ \mathcal{J}\\
		&s.t. \quad k_{n}\ge0, k_{n}^{(\tau)}\ge 0, k_n+k_n^{(\tau)}>0, n\in\{0,\ldots,m\},\\
		&\phantom{s.t. \quad} {k_\psi}>0, \bar{\sigma}>0, \underline{\sigma}\ge 0, \bar{\sigma}-\underline{\sigma}>0, \\
		&\phantom{s.t. \quad} \mu_2(\bar{T}\bar{A}_{ii}\bar{T}^{-1})\le -\bar{\sigma},  \\
  &\phantom{s.t. \quad}
  \sum_{j\in\sN_i}\norm{\bar{T}\bar{B}_{ij}(t)\bar{T}^{-1}}_2+\norm{\bar{T}\bar{B}_{ii}(t)\bar{T}^{-1}}_2\le \underline{\sigma},
	\end{split}
\end{align}
where $\bar{A}_{ii}$, $\bar{B}_{ii}, \bar{B}_{ij}$ are given by \Cref{equ: matrix} -- the explicit expressions for these matrices together with the expression of the transformation matrix $\bar{T}$ are given, for completeness, in \Cref{equ: matrix_general} where $\bar{g}_n:={k_\psi}k_n^{(\tau)}$, $n\in \{0,\ldots,m\}$.  
\begin{figure*}[b!]
\hrulefill
\begin{equation}\label{equ: matrix_general}
\begin{split}
    &\bar{T}=\left[\begin{matrix} I_n & \alpha_1 I_n & 0_n & \cdots & 0_n\\
     			0_n & I_n & \alpha_2 I_n & \cdots & 0_n\\ 						\vdots & \vdots & \vdots & \ddots & \vdots\\
     			0_n & 0_n & 0_n & \cdots & \alpha_m I_n\\
      			0_n& 0_n & 0_n & \cdots & I_n 	\end{matrix}\right], 
         \bar{B}_{ii}(t)=\abs{\sN_i}\frac{\partial \tanh(\eta_j-\eta_i-\delta_{ji}^\ast)}{\partial \eta_i}\left[\begin{matrix} \bar{g}_0 I_n & 					0_n & \cdots & 0_n\\
      			\vdots & \vdots & \ddots & \vdots\\
      			\bar{g}_m I_n & 0_n & \cdots & 0_n  \end{matrix}\right]\\
    &\bar{A}_{ii}=\left[\begin{matrix} -k_{0}I_n & I_n & 0_n & \cdots & 0_n\\
     			-k_{1}I_n & 0_n & I_n & \cdots & 0_n\\ 						\vdots & \vdots & \vdots & \ddots & \vdots\\
     			-k_{m-1}I_n & 0_n & 0_n & \cdots & I_n\\
      			-k_{m}I_n & 0_n & 0_n & \cdots & 0_n 	\end{matrix}\right],  
    \bar{B}_{ij}(t)=\left[\begin{matrix} \bar{g}_0 I_n & 					0_n & \cdots & 0_n\\
      			\vdots & \vdots & \ddots & \vdots\\
      			\bar{g}_m I_n & 0_n & \cdots & 0_n  \end{matrix}\right]\frac{\partial \tanh(\eta_j-\eta_i-\delta_{ji}^\ast)}{\partial \eta_j}
\end{split}
\end{equation}
\end{figure*}
The decision variables are $\bar{\xi}:=[k_{0},\ldots, k_{m}, k_{0}^{(\tau)}, \ldots, k_{m}^{(\tau)}, {k_\psi}, \allowbreak \underline{\sigma}, \bar{\sigma}, \alpha_{1}, \ldots, \alpha_{m}]$. As in Appendix \ref{sec: application}, the cost function can be chosen to maximize the upper bound of the inter-robot coupling functions (see the discussion in Appendix \ref{sec: application}). Then, following the same steps used to obtain \Cref{equ: optimisation} the constraints in \Cref{equ: optimisation_general} can be recast using LMIs and this yields the analogous of \Cref{equ: cvx_optimisation}:
\begin{align}\label{equ: optimisation_general_2}
	\begin{split}
		&\min_{\xi} \ \mathcal{J}\\
		&s.t. \quad k_{n}\ge0, \bar{g}_{n}\ge 0, k_n+\bar{g}_n>0, n\in\{0,\ldots,m\}, \\
		&\phantom{s.t. \quad} \bar{\sigma}>0, \underline{\sigma}\ge 0, \bar{\sigma}-\underline{\sigma}>0, \\
            &\phantom{s.t. \quad}[\bar{T}\bar{A}_{ii}\bar{T}^{-1}]_s\le -\bar{\sigma}I_{(m+1)n},\\
		&\phantom{s.t. \quad}\left[\begin{matrix}\frac{\underline{\sigma}}{2\bar{N}}I_{(m+1)n} & (\bar{T}\tilde{B}_{ij}\bar{T}^{-1})\T \\ \bar{T}\tilde{B}_{ij}\bar{T}^{-1} & \frac{\underline{\sigma}}{2\bar{N}}I_{(m+1)n}\end{matrix}\right]\succeq 0, \\
  &\phantom{s.t. \quad} \left[\begin{matrix}\frac{\underline{\sigma}}{2}I_{(m+1)n} & (\bar{T}\tilde{B}_{ii}\bar{T}^{-1})\T \\ \bar{T}\tilde{B}_{ii}\bar{T}^{-1} & \frac{\underline{\sigma}}{2}I_{(m+1)n}\end{matrix}\right]\succeq 0.
	\end{split}
\end{align}
where $\xi=[k_{0},\ldots, k_{m}, \bar{g}_0,\allowbreak \ldots,\allowbreak \bar{g}_m,  \underline{\sigma}, \bar{\sigma}]$ and 
\begin{align*}
\begin{split}
    \tilde{B}_{ii}&=-\bar{N}\left[\begin{matrix} \bar{g}_0 I_n & 					0_n & \cdots & 0_n\\
      			\vdots & \vdots & \ddots & \vdots\\
      			\bar{g}_m I_n & 0_n & \cdots & 0_n  \end{matrix}\right], \\
    \tilde{B}_{ij}&=\left[\begin{matrix} \bar{g}_0 I_n & 					0_n & \cdots & 0_n\\
      			\vdots & \vdots & \ddots & \vdots\\
      			\bar{g}_m I_n & 0_n & \cdots & 0_n  \end{matrix}\right]. 
\end{split}
\end{align*}

\section*{Acknowledgments}
The authors would like to thank Prof. Francesco Bullo for the insightful discussions on an early version of these results. These inputs helped us to improve the quality of our paper. {The authors also wish to thank the AE and five anonymous reviewers who made several helpful comments and suggestions. These inputs led to improvements over the originally submitted manuscript.}

\bibliographystyle{IEEEtran}
%\bibliography{Reference_network}

\begin{thebibliography}{10}
\providecommand{\url}[1]{#1}
\csname url@samestyle\endcsname
\providecommand{\newblock}{\relax}
\providecommand{\bibinfo}[2]{#2}
\providecommand{\BIBentrySTDinterwordspacing}{\spaceskip=0pt\relax}
\providecommand{\BIBentryALTinterwordstretchfactor}{4}
\providecommand{\BIBentryALTinterwordspacing}{\spaceskip=\fontdimen2\font plus
\BIBentryALTinterwordstretchfactor\fontdimen3\font minus
  \fontdimen4\font\relax}
\providecommand{\BIBforeignlanguage}[2]{{%
\expandafter\ifx\csname l@#1\endcsname\relax
\typeout{** WARNING: IEEEtran.bst: No hyphenation pattern has been}%
\typeout{** loaded for the language `#1'. Using the pattern for}%
\typeout{** the default language instead.}%
\else
\language=\csname l@#1\endcsname
\fi
#2}}
\providecommand{\BIBdecl}{\relax}
\BIBdecl

\bibitem{https://doi.org/10.48550/arxiv.2202.07638}
S.~Xie and G.~Russo, ``On the design of scalable networks rejecting first order
  disturbances,'' \emph{IFAC-PapersOnLine}, vol.~55, no.~13, pp. 216--221,
  2022.

\bibitem{MONTEIL2019198}
J.~Monteil, G.~Russo, and R.~Shorten, ``On {$\mathcal{L}_\infty$} string
  stability of nonlinear bidirectional asymmetric heterogeneous platoon
  systems,'' \emph{Automatica}, vol. 105, pp. 198--205, 2019.

\bibitem{STUDLI2017157}
S.~Stüdli, M.~Seron, and R.~Middleton, ``From vehicular platoons to general
  networked systems: String stability and related concepts,'' \emph{Annual
  Reviews in Control}, vol.~44, pp. 157--172, 2017.

\bibitem{mesh}
A.~Pant, P.~Seiler, and K.~Hedrick, ``Mesh stability of look-ahead
  interconnected systems,'' \emph{IEEE Transactions on Automatic Control},
  vol.~3, pp. 403 -- 407, 2002.

\bibitem{1303690}
H.~Tanner, G.~Pappas, and V.~Kumar, ``Leader-to-formation stability,''
  \emph{IEEE Transactions on Robotics and Automation}, vol.~20, no.~3, pp.
  443--455, 2004.

\bibitem{8370724}
B.~Besselink and S.~Knorn, ``Scalable input-to-state stability for performance
  analysis of large-scale networks,'' \emph{IEEE Control Systems Letters},
  vol.~2, no.~3, pp. 507--512, 2018.

\bibitem{10156367}
M.~Mirabilio and A.~Iovine, ``Scalable stability of nonlinear interconnected
  systems in case of amplifying perturbations,'' in \emph{2023 American Control
  Conference (ACC)}, 2023, pp. 3584--3589.

\bibitem{tegling2023scale}
E.~Tegling, B.~Bamieh, and H.~Sandberg, ``Scale fragilities in localized
  consensus dynamics,'' \emph{Automatica}, vol. 153, p. 111046, 2023.

\bibitem{9353260}
S.~Xie, G.~Russo, and R.~H. Middleton, ``Scalability in nonlinear network
  systems affected by delays and disturbances,'' \emph{IEEE Transactions on
  Control of Network Systems}, vol.~8, no.~3, pp. 1128--1138, 2021.

\bibitem{5457987}
K.-S. Kim, K.-H. Rew, and S.~Kim, ``Disturbance observer for estimating higher
  order disturbances in time series expansion,'' \emph{IEEE Transactions on
  Automatic Control}, vol.~55, no.~8, pp. 1905--1911, 2010.

\bibitem{6425973}
G.~Park, Y.~Joo, H.~Shim, and J.~Back, ``Rejection of polynomial-in-time
  disturbances via disturbance observer with guaranteed robust stability,'' in
  \emph{51st IEEE Conference on Decision and Control}, 2012, pp. 949--954.

\bibitem{knorn2014passivity}
S.~Knorn, A.~Donaire, J.~C. Ag{\"u}ero, and R.~H. Middleton, ``Passivity-based
  control for multi-vehicle systems subject to string constraints,''
  \emph{Automatica}, vol.~50, no.~12, pp. 3224--3230, 2014.

\bibitem{SILVA2021109542}
G.~F. Silva, A.~Donaire, A.~McFadyen, and J.~J. Ford, ``String stable integral
  control design for vehicle platoons with disturbances,'' \emph{Automatica},
  vol. 127, p. 109542, 2021.

\bibitem{di2015design}
M.~di~Bernardo, P.~Falcone, A.~Salvi, and S.~Santini, ``Design, analysis, and
  experimental validation of a distributed protocol for platooning in the
  presence of time-varying heterogeneous delays,'' \emph{IEEE Transactions on
  Control Systems Technology}, vol.~24, no.~2, pp. 413--427, 2015.

\bibitem{li2020longitudinal}
Y.~Li, Z.~Zhong, Y.~Song, Q.~Sun, H.~Sun, S.~Hu, and Y.~Wang, ``Longitudinal
  platoon control of connected vehicles: Analysis and verification,''
  \emph{IEEE Transactions on Intelligent Transportation Systems}, vol.~23,
  no.~5, pp. 4225--4235, 2020.

\bibitem{hu2023spontaneous}
B.-B. Hu, H.-T. Zhang, W.~Yao, J.~Ding, and M.~Cao, ``Spontaneous-ordering
  platoon control for multirobot path navigation using guiding vector fields,''
  \emph{IEEE Transactions on Robotics}, vol.~39, no.~4, pp. 2654--2668, 2023.

\bibitem{LOHMILLER1998683}
W.~Lohmiller and J.-J. Slotine, ``On contraction analysis for non-linear
  systems,'' \emph{Automatica}, vol.~34, no.~6, pp. 683--696, 1998.

\bibitem{centorrino2023euclidean}
V.~Centorrino, A.~Gokhale, A.~Davydov, G.~Russo, and F.~Bullo, ``Euclidean
  contractivity of neural networks with symmetric weights,'' \emph{IEEE Control
  Systems Letters}, vol.~7, pp. 1724--1729, 2023.

\bibitem{aminzare2014contraction}
Z.~Aminzare and E.~D. Sontag, ``Contraction methods for nonlinear systems: A
  brief introduction and some open problems,'' in \emph{53rd IEEE Conference on
  Decision and Control}.\hskip 1em plus 0.5em minus 0.4em\relax IEEE, 2014, pp.
  3835--3847.

\bibitem{tsukamoto2021contraction}
H.~Tsukamoto, S.-J. Chung, and J.-J.~E. Slotine, ``Contraction theory for
  nonlinear stability analysis and learning-based control: A tutorial
  overview,'' \emph{Annual Reviews in Control}, vol.~52, pp. 135--169, 2021.

\bibitem{1618853}
W.~Wang and J.-J. Slotine, ``Contraction analysis of time-delayed
  communications and group cooperation,'' \emph{IEEE Transactions on Automatic
  Control}, vol.~51, no.~4, pp. 712--717, 2006.

\bibitem{8561231}
H.~S. Shiromoto, M.~Revay, and I.~R. Manchester, ``Distributed nonlinear
  control design using separable control contraction metrics,'' \emph{IEEE
  Transactions on Control of Network Systems}, vol.~6, no.~4, pp. 1281--1290,
  2019.

\bibitem{COOGAN2019349}
S.~Coogan, ``A contractive approach to separable {Lyapunov} functions for
  monotone systems,'' \emph{Automatica}, vol. 106, pp. 349--357, 2019.

\bibitem{9247460}
J.~W. Simpson-Porco, ``Analysis and synthesis of low-gain integral controllers
  for nonlinear systems,'' \emph{IEEE Transactions on Automatic Control},
  vol.~66, no.~9, pp. 4148--4159, 2021.

\bibitem{9627575}
M.~Giaccagli, D.~Astolfi, V.~Andrieu, and L.~Marconi, ``Sufficient conditions
  for global integral action via incremental forwarding for input-affine
  nonlinear systems,'' \emph{IEEE Transactions on Automatic Control}, vol.~67,
  no.~12, pp. 6537--6551, 2021.

\bibitem{entrainment_2010}
G.~Russo, M.~di~Bernardo, and E.~D. Sontag, ``Global entrainment of
  transcriptional systems to periodic inputs,'' \emph{PLoS computational
  biology}, vol.~6, no.~4, p. e1000739, 2010.

\bibitem{Marg_22}
R.~Ofir, F.~Bullo, and M.~Margaliot, ``Minimum effort decentralized control
  design for contracting network systems,'' \emph{IEEE Control Systems
  Letters}, vol.~6, pp. 2731--2736, 2022.

\bibitem{jafarpour2021robust}
S.~Jafarpour, A.~Davydov, A.~Proskurnikov, and F.~Bullo, ``Robust implicit
  networks via {non-Euclidean} contractions,'' \emph{Advances in Neural
  Information Processing Systems}, vol.~34, pp. 9857--9868, 2021.

\bibitem{1531-3492_2021188}
G.~Russo and F.~Wirth, ``Matrix measures, stability and contraction theory for
  dynamical systems on time scales,'' \emph{Discrete and Continuous Dynamical
  Systems - B}, vol.~27, no.~6, pp. 3345--3374, 2022.

\bibitem{https://doi.org/10.48550/arxiv.2209.08993}
S.~Xie and G.~Russo, ``On the design of multiplex control to reject
  disturbances in nonlinear network systems affected by heterogeneous delays,''
  in \emph{2023 American Control Conference (ACC)}.\hskip 1em plus 0.5em minus
  0.4em\relax IEEE, 2023, pp. 240--245.

\bibitem{5717887}
G.~Russo, M.~di~Bernardo, and E.~D. Sontag, ``Stability of networked systems: A
  multi-scale approach using contraction,'' in \emph{49th IEEE Conference on
  Decision and Control}, 2010, pp. 6559--6564.

\bibitem{WEN2008169}
L.~Wen, Y.~Yu, and W.~Wang, ``Generalized {Halanay} inequalities for
  dissipativity of volterra functional differential equations,'' \emph{Journal
  of Mathematical Analysis and Applications}, vol. 347, no.~1, pp. 169--178,
  2008.

\bibitem{6740883}
S.~Sridhar and M.~Govindarasu, ``Model-based attack detection and mitigation
  for automatic generation control,'' \emph{IEEE Transactions on Smart Grid},
  vol.~5, no.~2, pp. 580--591, 2014.

\bibitem{fridman2010refined}
E.~Fridman, ``A refined input delay approach to sampled-data control,''
  \emph{Automatica}, vol.~46, no.~2, pp. 421--427, 2010.

\bibitem{9462542}
E.~Abolfazli, B.~Besselink, and T.~Charalambous, ``On time headway selection in
  platoons under the {MPF} topology in the presence of communication delays,''
  \emph{IEEE Transactions on Intelligent Transportation Systems}, vol.~23,
  no.~7, pp. 8881--8894, 2022.

\bibitem{doi:10.1137/110832574}
W.~Qiao and R.~Sipahi, ``A linear time-invariant consensus dynamics with
  homogeneous delays: Analytical study and synthesis of rightmost
  eigenvalues,'' \emph{SIAM Journal on Control and Optimization}, vol.~51,
  no.~5, pp. 3971--3992, 2013.

\bibitem{guo2018event}
Z.~Guo, S.~Gong, S.~Wen, and T.~Huang, ``Event-based synchronization control
  for memristive neural networks with time-varying delay,'' \emph{IEEE
  Transactions on cybernetics}, vol.~49, no.~9, pp. 3268--3277, 2018.

\bibitem{BURBANOLOMBANA2016310}
D.~A. {Burbano Lombana} and M.~{di Bernardo}, ``Multiplex {PI} control for
  consensus in networks of heterogeneous linear agents,'' \emph{Automatica},
  vol.~67, pp. 310--320, 2016.

\bibitem{wilson2022robotarium}
S.~Wilson and M.~Egerstedt, ``The robotarium: A remotely-accessible,
  multi-robot testbed for control research and education,'' \emph{IEEE Open
  Journal of Control Systems}, vol.~2, pp. 12--23, 2022.

\bibitem{1261347}
J.~Lawton, R.~Beard, and B.~Young, ``A decentralized approach to formation
  maneuvers,'' \emph{IEEE Transactions on Robotics and Automation}, vol.~19,
  no.~6, pp. 933--941, 2003.

\bibitem{5979747}
D.~Panagou and K.~J. Kyriakopoulos, ``Switching control approach for the robust
  practical stabilization of a unicycle-like marine vehicle under non-vanishing
  perturbations,'' in \emph{2011 IEEE International Conference on Robotics and
  Automation}, 2011, pp. 1525--1530.

\bibitem{li2013leader}
W.~Li, Z.~Chen, and Z.~Liu, ``Leader-following formation control for
  second-order multiagent systems with time-varying delay and nonlinear
  dynamics,'' \emph{Nonlinear Dynamics}, vol.~72, no.~4, pp. 803--812, 2013.

\bibitem{siljak2011decentralized}
D.~D. Siljak, \emph{Decentralized control of complex systems}.\hskip 1em plus
  0.5em minus 0.4em\relax Courier Corporation, 2011.

\bibitem{chen2021mean}
G.~Chen, C.~Fan, J.~Sun, and J.~Xia, ``Mean square exponential stability
  analysis for {It{\^o}} stochastic systems with aperiodic sampling and
  multiple time-delays,'' \emph{IEEE Transactions on Automatic Control},
  vol.~67, no.~5, pp. 2473--2480, 2021.

\bibitem{chen2023controller}
G.~Chen, G.~Du, J.~Xia, X.~Xie, and J.~H. Park, ``Controller synthesis of
  aperiodic sampled-data networked control system with application to
  interleaved flyback module integrated converter,'' \emph{IEEE Transactions on
  Circuits and Systems I: Regular Papers}, vol.~70, no.~11, pp. 4570--4580,
  2023.

\bibitem{yao2021stochastic}
L.~Yao, Z.~Wang, X.~Huang, Y.~Li, Q.~Ma, and H.~Shen, ``Stochastic sampled-data
  exponential synchronization of {Markovian} jump neural networks with
  time-varying delays,'' \emph{IEEE Transactions on Neural Networks and
  Learning Systems}, vol.~34, no.~2, pp. 909--920, 2023.

\bibitem{1083507}
C.~Desoer and H.~Haneda, ``The measure of a matrix as a tool to analyze
  computer algorithms for circuit analysis,'' \emph{IEEE Transactions on
  Circuit Theory}, vol.~19, no.~5, pp. 480--486, 1972.

\bibitem{boyd2004convex}
S.~Boyd and L.~Vandenberghe, \emph{Convex optimization}.\hskip 1em plus 0.5em
  minus 0.4em\relax Cambridge university press, 2004.

\end{thebibliography}

% Generated by IEEEtran.bst, version: 1.14 (2015/08/26)

\begin{IEEEbiography}[{\includegraphics[width=1in,height=1.25in,clip,trim=12cm 0cm 6cm 4cm]{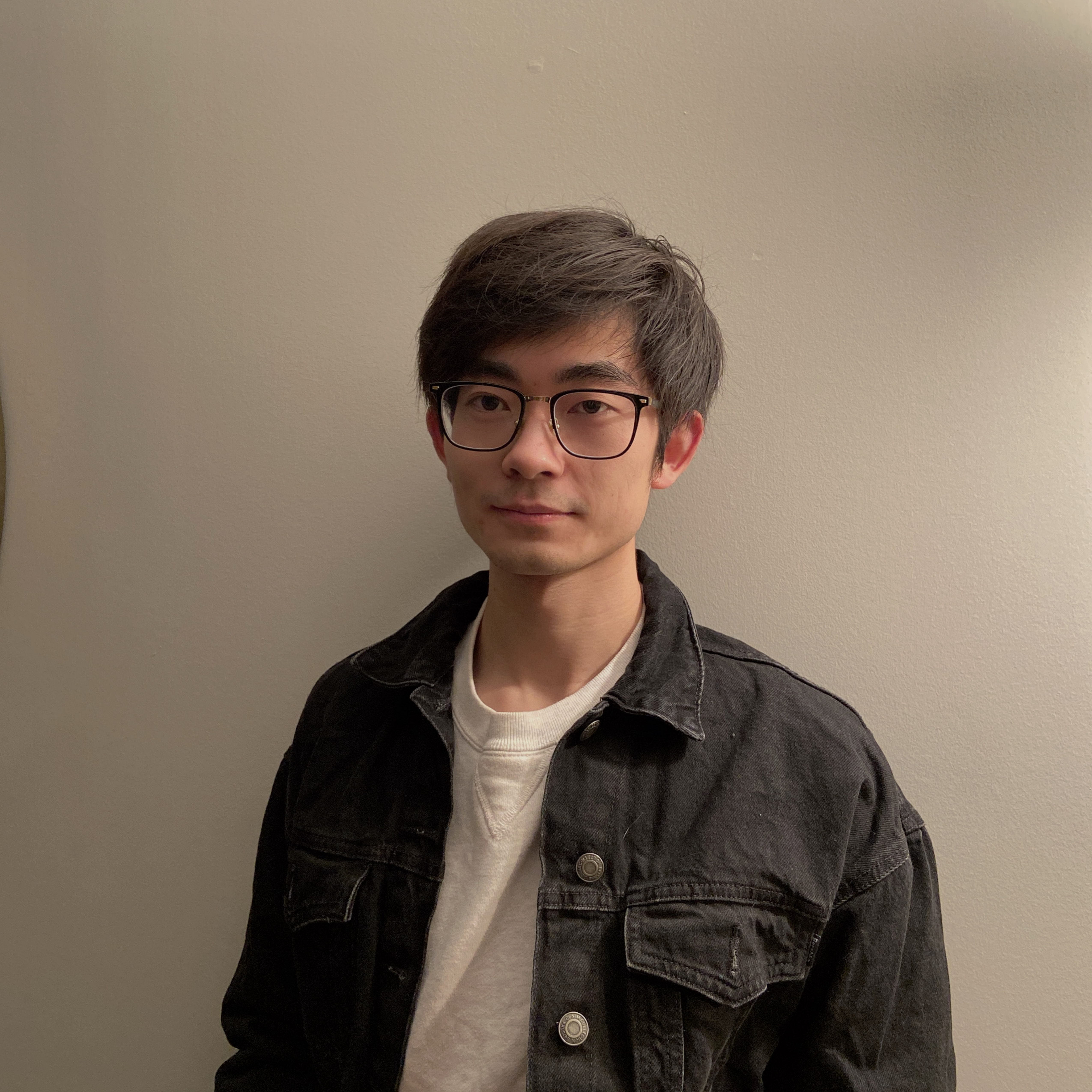}}]{Shihao Xie}{\space} received his Ph.D. degree in Electrical and Electronic Engineering from University College Dublin, Dublin, Ireland, in 2023. He is currently a postdoctoral research fellow with the Department of Electrical Engineering and Information Technologies at University of Naples Federico II, Naples, Italy. His research interests is mainly with the control of network systems with applications to e.g. robotic network, neural network.
\end{IEEEbiography}

\begin{IEEEbiography}[{\includegraphics[width=1in,height=1.25in,clip,trim=2.5cm 6.5cm 2.5cm 2.5cm]{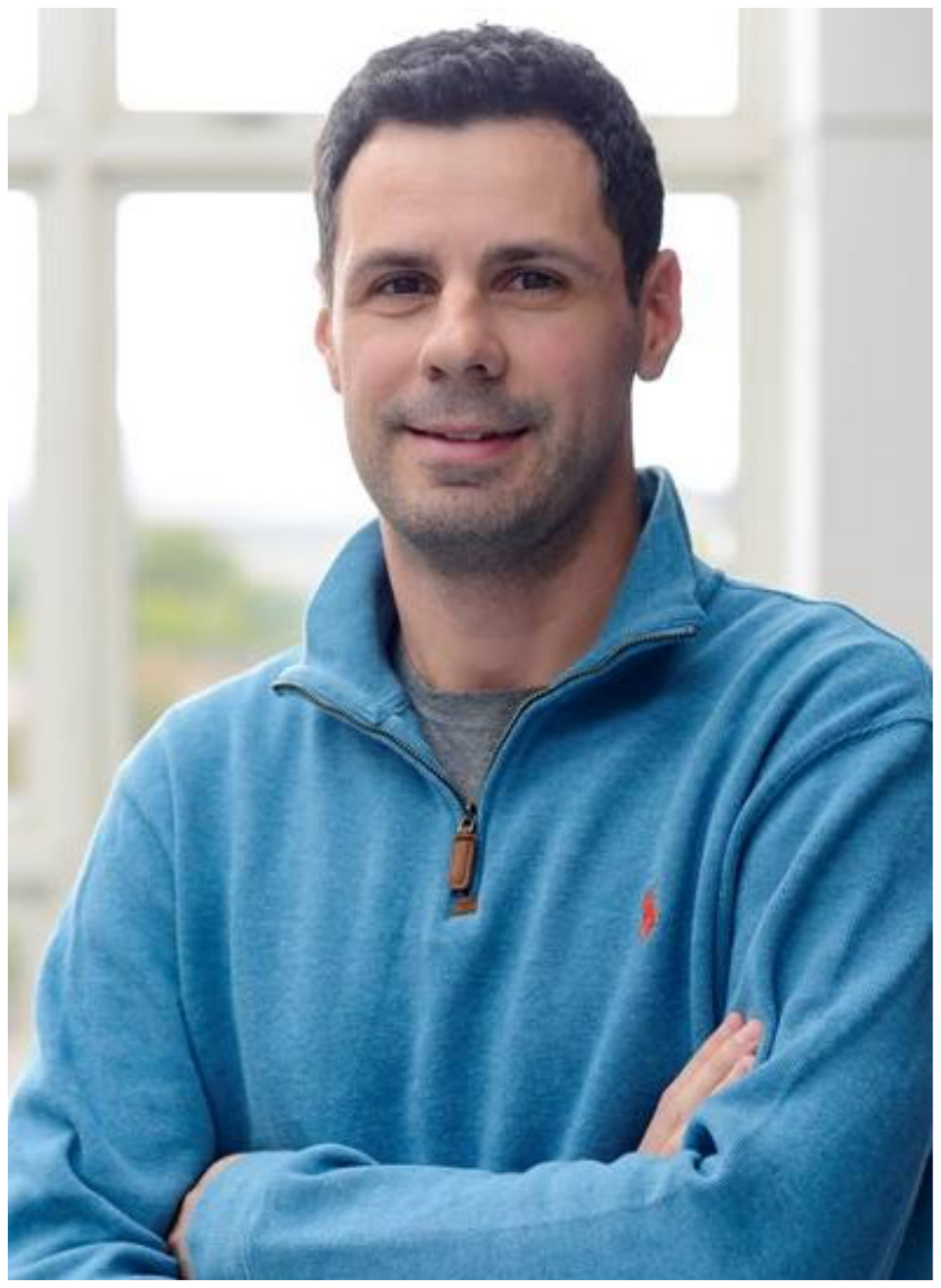}}]{Giovanni Russo}{\space}(Senior Member, IEEE) is an Associate Professor of Automatic Control at the University of Salerno, Italy. He was previously with the University of Naples Federico II (Ph.D. in 2010), Italy, Ansaldo STS (System Engineer/Integrator of the Honolulu Rail Transit Project, USA in 2012-2015), IBM Research Ireland (Research Staff Member in Optimization, Control and Decision Science from 2015 to 2018) and University College Dublin, Ireland (in 2018-2020). His research interests include contraction theory, analysis/control of nonlinear and complex systems, data-driven sequential decision-making under uncertainty and control in the space of densities. Dr. Russo has served as Associate Editor for the IEEE Transactions on Circuits and Systems I: regular papers (2016-2019) and the IEEE Transactions on Control of Network Systems (2017-2023). Since January 2024, Dr. Russo is serving as Senior Editor for the IEEE Transactions on Control of Network Systems. Personal page: \url{https://tinyurl.com/2p8zfpme}.
\end{IEEEbiography}

\end{document}